\documentclass{article}
\usepackage{iclr2027_conference,times}

\usepackage[utf8]{inputenc}
\usepackage[T1]{fontenc}
\usepackage{url}
\usepackage{booktabs,tabularx,multirow,longtable}
\usepackage{amsmath,amssymb,amsfonts}
\usepackage{nicefrac}
\usepackage{microtype}
\usepackage{graphicx,xcolor}
\usepackage{algpseudocode}
\usepackage{pdflscape}
\usepackage{hyperref}
\usepackage{enumitem}
\usepackage{wrapfig}
\usepackage[linesnumbered,ruled,lined]{algorithm2e}
\usepackage{marvosym}
\usepackage{fontawesome5}
\usepackage{dsfont}

\hypersetup{
    colorlinks=true,
    citecolor=teal,
    linkcolor=red,
    filecolor=magenta,
    urlcolor=magenta}

\newcommand{\method}{\textsc{BridgeCast}}
\newcommand{\mat}[1]{{\bf #1}}   
\newcommand{\red}[1]{\textcolor{red}{#1}}
\newcommand{\blue}[1]{\textcolor{blue}{#1}}

\SetCommentSty{mycommfont}

\title{\raisebox{-0.15em}{\includegraphics[height=0.9em]{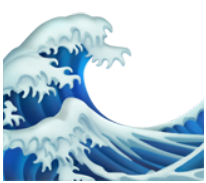}} \method: Bridging Ocean Wave Forecasts to Reanalysis via Flow Matching with Exogenous Variables}
\author{Siyu Gan\textsuperscript{1}, Dongsheng Luo\textsuperscript{2}, Kunxiaojia Yuan\textsuperscript{3}, Dongjin Song\textsuperscript{4}, Jingchao Ni\textsuperscript{3}\textsuperscript{\Letter}\\
\textsuperscript{1}Johns Hopkins University, \textsuperscript{2}Singapore Management University,\\
\textsuperscript{3}University of Houston, \textsuperscript{4}University of Connecticut\\
\textsuperscript{1}sgan5@jh.edu, \textsuperscript{2}dsluo@smu.edu.sg, \textsuperscript{3}\{kyuan, jni7\}@uh.edu, \textsuperscript{4}dongjin.song@uconn.edu
\vspace{-0.6cm}}

\iclrfinalcopy 

\begin{document}
\maketitle

\begin{abstract}
Ocean wave forecasting is essential for maritime safety, offshore operations, and coastal resilience, yet remains challenging due to systematic biases in physics-based models. Physical models, while widely used, rely on approximations and parameterizations that limit their accuracy under complex ocean-atmosphere conditions. To enhance ocean wave forecasting, we propose \method, within a physics-AI hybrid framework for bias correction. \method\ is a probabilistic model based on conditional flow matching (CFM) that learns to transform physical model forecasts into reanalysis-like fields. It treats physical forecasts as corrupted observations and employs a continuous-time generative process to bridge their distribution toward that of reanalysis data. \method\ is parameterized by a Transformer-based architecture that enables spatiotemporal modeling, incorporation of exogenous atmospheric variables, and flexible inference via both ordinary and stochastic differential equation formulations. Extensive experiments on real-world datasets demonstrate that \method\ consistently outperforms state-of-the-art baselines across regions and forecast lead times.
\end{abstract}

\section{Introduction}


\vspace{-0.2cm}

\begin{figure}
\centering
\includegraphics[width=0.93\linewidth]{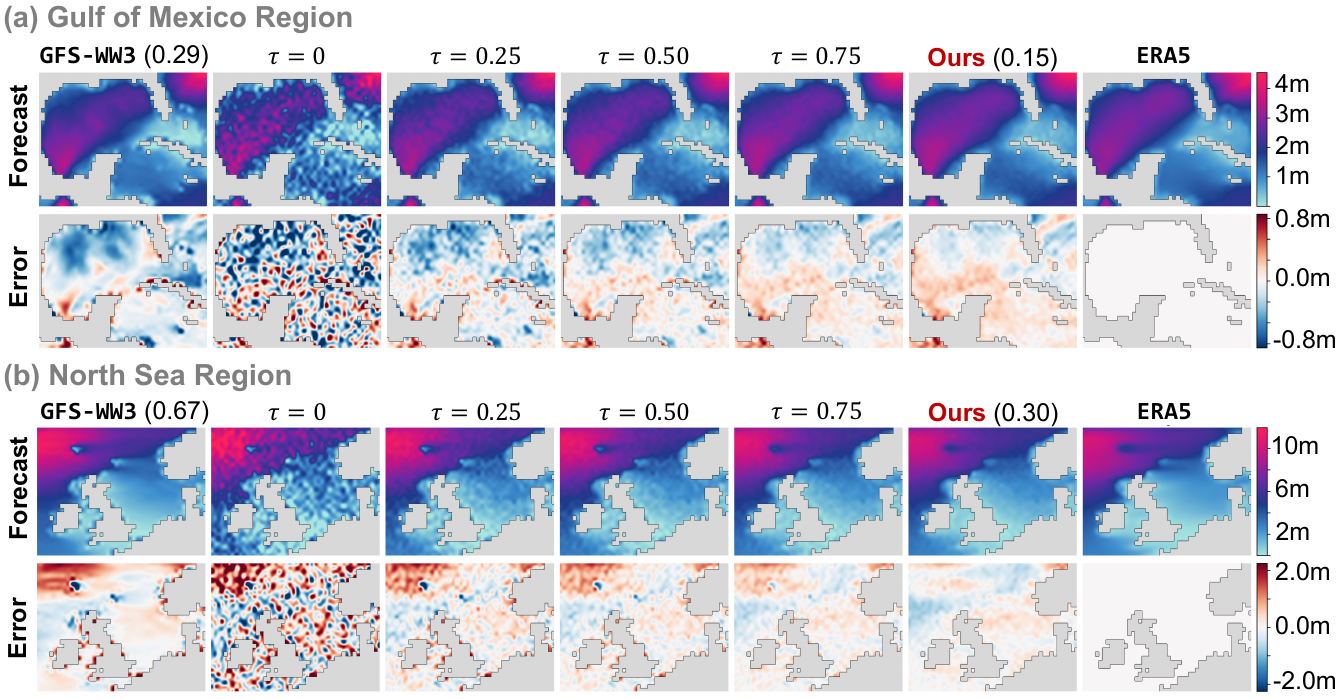}
\caption{An illustration of \method's process of bridging \texttt{GFS-WW3}'s forecasts to the reanalysis data of \texttt{ERA5} in (a) Gulf of Mexico, and (b) North Sea. The numbers in brackets are RMSEs. $\tau$ is 
flow time. ``Ours'' is \method, 
which resembles \texttt{ERA5} with less errors than \texttt{GFS-WW3}.}\label{fig.example}
\end{figure}

Ocean wave forecasting is critical for maritime safety, offshore operations, energy infrastructure, hazard mitigation, and coastal resilience \citep{sun2022deep,zhang2024deep,fang2026waveuformer}. At present, operational wave forecasts are predominantly generated by physics-based numerical models, such as the spectral wave model \texttt{WaveWatch-III} (\texttt{WW3}) \citep{tolman2009ww3}, which is often forced by atmospheric forecasts from the Global Forecast System (\texttt{GFS}) operated by the National Centers for Environmental Prediction (NCEP), and the \texttt{WaveModel} (\texttt{WAM}) \citep{ecmwf-ifs-wam} used within the Integrated Forecasting System (\texttt{IFS}) operated by the European Centre for Medium-Range Weather Forecasts (ECMWF). These models simulate the evolution of the wave energy spectrum through governing equations that account for wind forcing, nonlinear wave-wave interactions, and dissipation processes. While grounded in well-established physical principles and widely used in practice, they inevitably rely on approximations, parameterizations, and imperfect boundary conditions, leading to systematic biases and reduced accuracy under complex or rapidly changing ocean-atmosphere conditions such as storms. For example, Fig. \ref{fig.example} (left column) shows the forecast errors of \texttt{GFS-WW3} in the Gulf of Mexico and North Sea in 2025 (lead-time: 30 h), varying within $\pm$0.8m (RMSE: 0.29) and $\pm$2.0m (RMSE: 0.67). 
These errors are substantial and may lead to critical misjudgments, posing safety risks for maritime and coastal operations.


To correct the biases of the physical forecasters, we introduce a physics-AI hybrid framework that improves their outputs using a novel deep learning model, \method. \method\ is a probabilistic predictor designed to bridge the distribution of physical model forecasts to that of reanalysis data using conditional flow matching (CFM) \citep{lipman2023flow,tong2024improving}. Reanalysis data, such as those provided by the ECMWF Reanalysis 5th Generation (\texttt{ERA5}) \citep{hersbach2020era5}, are generated through data assimilation that integrates observational data 
and are widely used as a proxy for ground truth in evaluating oceanic and atmospheric forecasts due to their consistency and reliability \citep{hersbach2020era5,sharmar2020validation,zhou2026comparing}. \method\ treats physical model forecasts as ``corrupted images'', which serve as conditions for a generative model based on continuous-time stochastic processes, enabling the generation of more accurate reanalysis-like outputs. Fig. \ref{fig.example} illustrates the transport from \texttt{GFS-WW3} forecasts to the \texttt{ERA5} distribution by \method, where $\tau$ is the flow time,  ``Ours'' refers to bias-corrected forecasts, and \texttt{ERA5} represents the reanalysis data. \method\ progressively refines a perturbed forecast ($\tau=0$) and ultimately reduces forecasting errors, producing outputs that closely resemble reanalysis fields.


To our knowledge, \method\ is among the first to apply CFM for bias correction in physical models for ocean wave forecasting, and can serve as a plug-and-play module for existing operational systems ({\em e.g.}, \texttt{GFS-WW3}, \texttt{IFS-WAM}) within a physics-AI hybrid framework that inherits the advantages of physical modeling while mitigating their limitations through ``learning to bridge'' distributions. Additionally, \method\ is parameterized by a novel two-head Diffusion Transformer (\texttt{DiT} \citep{peebles2023dit}) encoder-decoder capable of (1) spatiotemporal modeling of the geoscientific data; and (2) incorporating of exogenous variables such as atmospheric forcing factors that have a critical impact on ocean dynamics. Moreover, \method\ supports inferences using either an ordinary differential equation (ODE) formulation for efficient computation or a stochastic differential equation (SDE) formulation for uncertainty-aware forecasting, providing flexibility depending on practical requirements. In summary, our contributions are as follows.
\begin{itemize}[topsep=0pt,leftmargin=*]
\setlength{\itemsep}{0.01cm}
\item We investigate a physics-AI hybrid framework and study the problem of correcting biases in physical models for ocean wave forecasting using deep learning-based CFM.
\item We propose \method, a novel CFM-based probabilistic model that bridges the distribution of forecasts from physical models and that of reanalysis data. \method\ adopts a Transformer-based architecture supporting spatiotemporal modeling, exogenous variables, and flexible inference.
\item We conduct comprehensive experiments using \texttt{GFS}, \texttt{IFS}, and \texttt{ERA5} data, and compare \method\ with state-of-the-art (SOTA) deterministic and probabilistic baselines, demonstrating its superior performance across different regions and lead times.
\end{itemize}

\section{Related Work}

\vspace{-0.1cm}

\noindent{\textbf{Bias Correction for Ocean Wave Forecasting}}. Bias correction and statistical post-processing have become increasingly important in ocean wave forecasting. Early studies focused on statistical and operational correction techniques \citep{vannitsem2021statistical}. Later works extended these ideas through spatiotemporal correction frameworks and climate-aware calibration methods \citep{ai2022spatiotemporal,lira2023role}. Recent advances increasingly employ deep learning models \citep{sun2022deep,zhang2024deep,fang2026waveuformer}. For example, \citep{sun2022deep} introduced a \texttt{BU-Net} architecture for correcting \texttt{GFS-WW3} forecasts; \citep{zhang2024deep} utilized \texttt{TrajGRU} \citep{shi2017deep} with wind-wave coupled inputs for real-time correction; \citep{lira2023role} applied \texttt{U-Net} and Transformer components to improve medium- and long-range wave forecasts. Related studies have further explored physics-informed corrections and upstream atmospheric adjustments \citep{meisner2023wave,peduto2025observation}. Despite these advances, most existing learning-based approaches \citep{sun2022deep,zhang2024deep,fang2026waveuformer} rely on deterministic regression objectives and do not explicitly model conditional forecast distributions or transport dynamics, leaving generative approaches such as flow matching underexplored for ocean wave forecast bias correction.

\noindent{\textbf{Earth System Predictions}}. Recent data-driven Earth models are mostly for weather and climate forecasting. Representative global forecasting systems include deterministic models \citep{lam2023learning,pathak2022fourcastnet} 
that aim to emulate or replace weather prediction at global scales. 
Precipitation nowcasting models 
\citep{shi2015convolutional,sonderby2020metnet,flowcast_ode} focus on rapid prediction of weather events. Foundation-style Earth system models such as \texttt{Earthformer} \citep{gao2022earthformer}, \texttt{ClimaX} \citep{nguyen2023climax}, and \texttt{NeuralGCM} \citep{kochkov2024neural} extend these capabilities toward multi-variable, multi-task modeling of the Earth system. Emerging probabilistic \citep{gong2024cascast,gao2025oneforecast}, latent-variable \citep{leinonen2023latent,wu2024earthfarseer}, and diffusion-based approaches \citep{yu2024diffcast,gao2023prediff} further explore generative formulations. However, these models are primarily designed for global weather or Earth system variables and are not tailored for ocean wave forecasting.

\noindent{\textbf{Flow-Based Models}}. Traditional continuous normalizing flows (CNFs) \citep{chen2018neural} model data via neural ODEs, but require costly numerical integration, which can be prohibitive for high-dimensional spatiotemporal applications. CFM \citep{lipman2023flow,tong2024improving} and Rectified Flow \citep{liu2023flow} methods address this limitation by directly regressing a time-dependent vector field against a conditional probability path, avoiding the need for solving ODEs during training. Unlike diffusion models, which rely on stochastic denoising processes that typically require many iterative sampling steps, CFM enables learning ODE trajectories that transport noise to data more efficiently. \texttt{FlowCast} \citep{flowcast_ode} is a recent example that applies CFM to precipitation nowcasting. More recently, stochastic interpolants \citep{albergo2023datadep,albergo2025stochastic} have been proposed as a unified framework connecting diffusion models and flow-based formulations. To the best of our knowledge, our work is the first to adapt CFM-based modeling to ocean wave forecast correction with explicit spatiotemporal modeling and exogenous variable conditioning.

\vspace{-0.1cm}


\section{The Proposed \method\ Model}\label{sec:method}

\vspace{-0.2cm}

\noindent{\textbf{Problem Statement}}. Bias correction for ocean wave forecasts is framed as an image sequence correction task with exogenous variables. In this task, a physical model generates a forecast of wave variables --- which are {\em 
endogenous variables} --- $\mathcal{X} = [\mat{X}_{1}, ..., \mat{X}_{T}]$ within a forecast horizon of length $T$, where $\mat{X}_{t}\in\mathbb{R}^{H\times W\times C}$ ($1\le t\le T$) is a forecasted image with $C$ channels ({\em i.e.}, variables) at time step $t$. Additionally, we are given observed 
atmospheric variables --- which are {\em exogenous variables} --- $\mathcal{Z}=[\mat{Z}_{1}, ..., \mat{Z}_{T'}]$ within a time window of length $T'$, where $\mat{Z}_{t'}\in\mathbb{R}^{H'\times W'\times C'}$ ($1\le t\le T'$) is an image with $C'$ channels at time step $t'$. It is noteworthy that $\mathcal{X}$ and $\mathcal{Z}$ can have different lengths and sizes. The goal is to predict wave variables within the physical model's forecasting horizon $\hat{\mathcal{Y}}=[\hat{\mat{Y}}_{1}, ..., \hat{\mat{Y}}_{T}]$ where $\hat{\mat{Y}}_{t}\in\mathbb{R}^{H\times W\times C}$, such that $\hat{\mathcal{Y}}$ is closer to the reanalysis data $\mathcal{Y}=[\mat{Y}_{1}, ..., \mat{Y}_{T}]$ than the physical model's raw output $\mathcal{X}$.


\noindent{\textbf{Preliminaries}}. \method\ is built upon 
CFM \citep{albergo2025si}, 
which defines a conditional distribution $p_{\tau}(x_{\tau}|x_{0}, x_{1})$, typically a Gaussian distribution with mean $(1-\tau)x_{0} + \tau x_{1}$ and time-dependent standard deviation $\sigma_{\tau}$, where $0\le\tau\le 1$. This distribution establishes a bridge that interpolates between a source data sample $x_{0}\sim p_{0}$ and a target data sample $x_{1}\sim p_{1}$. It basically defines a stochastic process $x_{\tau}=(1-\tau)x_{0} + \tau x_{1} + \sigma_{\tau}\epsilon$, where $\epsilon\sim \mathcal{N}(0, \mat{I})$, that can bridge two arbitrary distributions $p_{0}$ and $p_{1}$ \citep{albergo2025stochastic}. 
CFM trains a velocity field $v_{\boldsymbol{\theta}}(x, \tau)$ by regressing it against the velocity target $\frac{dx_{\tau}}{d\tau}$. 
Additionally, the 
framework trains 
a denoiser function $s_{\boldsymbol{\theta}}(x, \tau)$ by regressing it against the 
sampled Gaussian noise $\epsilon$. 
Next, we specify the proposed \method\ under the 
CFM framework ($\S$\ref{sec.formulation}), and introduce our parameterization of $v_{\boldsymbol{\theta}}(x, \tau)$ and $s_{\boldsymbol{\theta}}(x, \tau)$ ($\S$\ref{sec.architecture}).

\subsection{\method\ Formulation}\label{sec.formulation}

Given a physical model's forecast $\mathcal{X}$, we aim to bridge its underlying distribution with that of 
reanalysis data $\mathcal{Y}$ during model training. Following the CFM framework, we define a stochastic interpolant \citep{albergo2025stochastic}
\begin{equation}\label{eq.stochastic}
\begin{aligned}
\hat{\mathcal{Y}}_{\tau} = (1-\tau)\mathcal{X} + \tau \mathcal{Y} + \sigma_{\tau}\boldsymbol{\epsilon},~~~\text{where}~~~\boldsymbol{\epsilon}\sim\mathcal{N}(\mat{0}, \mat{I})
\end{aligned}
\end{equation}
which is a flow between $\mathcal{X}$ and $\mathcal{Y}$. Specifically, we use $\sigma_{\tau}=(1-\tau)\sigma$ where $\sigma$ is a hyperparameter so that the noise $\sigma_{\tau}\boldsymbol{\epsilon}$ gradually vanishes as $\tau$ moves from 0 to 1. Thus, Eq.~\eqref{eq.stochastic} can be rewritten as
\begin{equation}\label{eq.perturb}
\begin{aligned}
\hat{\mathcal{Y}}_{\tau} = (1-\tau)(\mathcal{X} + \sigma\boldsymbol{\epsilon}) + \tau \mathcal{Y},
\end{aligned}
\end{equation}


Eq.~\eqref{eq.perturb} can be interpreted as a flow between a perturbed source $\mathcal{X} + \sigma\boldsymbol{\epsilon}$ and the clean target $\mathcal{Y}$. 
This corresponds to a one-sided noisy interpolation \citep{albergo2025stochastic}. Empirically, it transports a smoothed version of the source sample that makes the training process more stable. 
Consequently, our conditional distribution $p_{\tau}(\hat{\mathcal{Y}}_{\tau} | \mathcal{X}, \mathcal{Y})$ is a Gaussian distribution $\mathcal{N}\big((1-\tau)\mathcal{X} + \tau \mathcal{Y}, (1-\tau)^{2}\sigma^{2}\mat{I}\big)$.

\noindent{\textbf{Training Objective}}. Starting from an SDE-based formulation that defines a stochastic probability path, we derive a hybrid training objective combining CFM and diffusion-style score matching under a unified interpolant framework. Using SDE explicitly equips \method\ with the capability of representing the probabilistic structures of ocean and atmospheric data including uncertainty and stochastic variability through an explicit noise process. In contrast, the counterpart ODE formulation provides a deterministic flow that represents the same marginal distributions but does not explicitly model intermediate stochasticity along trajectories.


Inspired by \citep{albergo2025si}, to approximate the decomposed velocity with a 
parameterized velocity field $v_{\boldsymbol{\theta}}(\cdot)$ and a parameterized 
denoiser function $s_{\boldsymbol{\theta}}(\cdot)$, we derive a joint training 
objective (the full derivation process is deferred to Appendix~\ref{app.theory})
\begin{equation}\label{eq.train_loss}
\begin{aligned}
\ell(\boldsymbol{\theta}) = \mathbb{E}_{\substack{(\mathcal{X}, \mathcal{Z}, \mathcal{Y})\sim\mathcal{D}_{\text{tr}},\\\tau\sim\mathcal{U}(0, 1),\boldsymbol{\epsilon}\sim\mathcal{N}(\mat{0}, \mat{I})}}[\|v_{\boldsymbol{\theta}}(\hat{\mathcal{Y}}_{\tau}, \mathcal{Z}, \tau) - (\mathcal{Y} - \mathcal{X}-\sigma\boldsymbol{\epsilon})\|^{2} + \|s_{\boldsymbol{\theta}}(\hat{\mathcal{Y}}_{\tau}, \mathcal{Z}, \tau) - \boldsymbol{\epsilon}\|^{2}]
\end{aligned}
\end{equation}
where $\boldsymbol{\epsilon}\sim\mathcal{N}(\mat{0}, \mat{I})$, and $\mathcal{D}_{\text{tr}}$ is the training dataset. 
The first term in Eq.~\eqref{eq.train_loss} is a flow-style regression of velocity field. The second term is a denoising score matching that empirically alleviates overfitting of model parameters $\boldsymbol{\theta}$ during training.




\subsection{\method\ Architecture}\label{sec.architecture}

\begin{figure}[!t]
\centering
\includegraphics[width=0.95\linewidth]{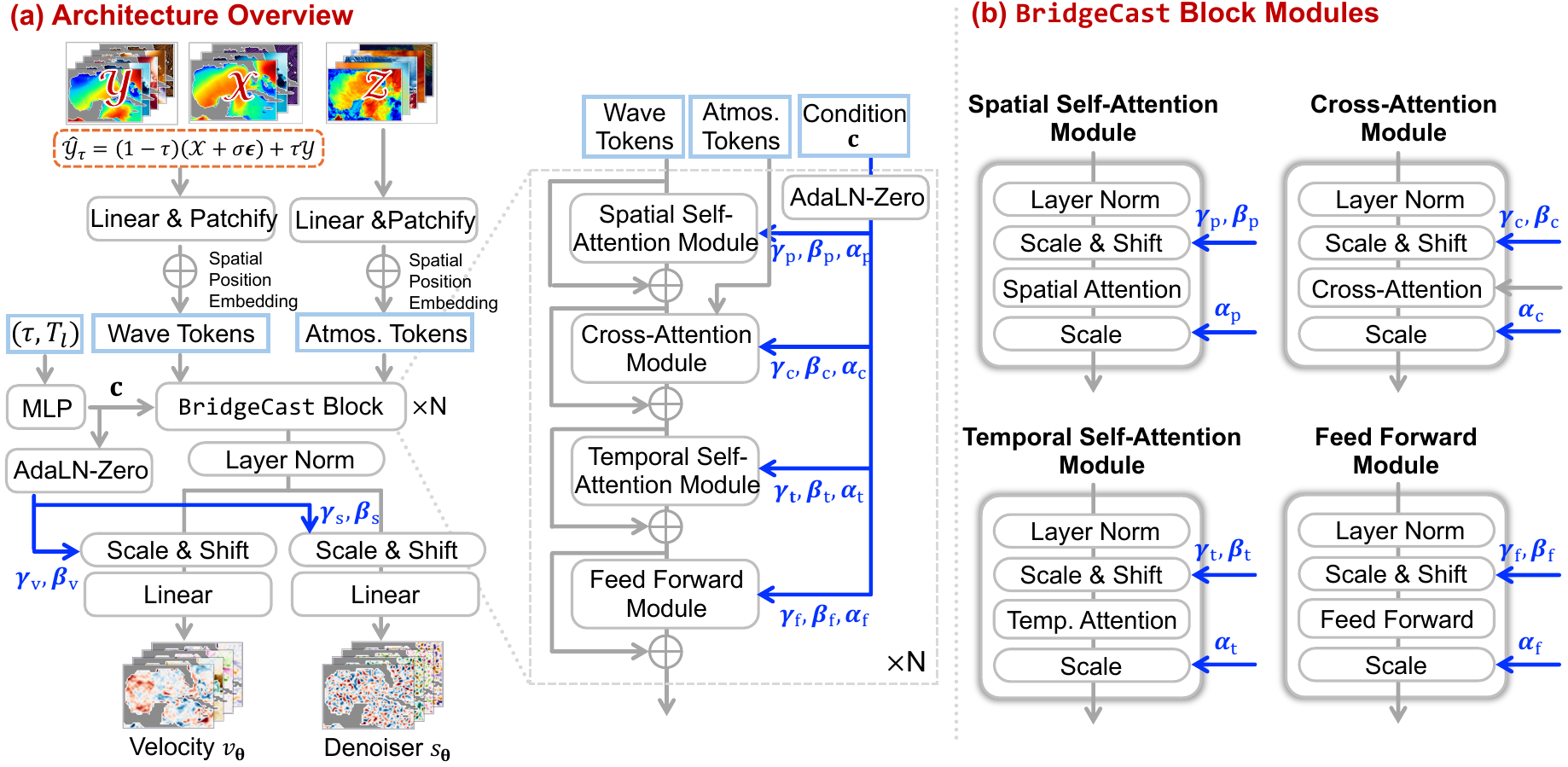}
\caption{An illustration of the \method\ Architecture. (a) is an overview of the architecture with a zoom-in of the \method\ block. (b) illustrates the four modules in the \method\ block. $\tau$ is the flow time. $T_{l}$ is the lead time. $\boldsymbol{\alpha}$'s, $\boldsymbol{\beta}$'s, $\boldsymbol{\gamma}$'s are scaling and shifting parameters.}\label{fig.architecture}
\vspace{-0.1cm}
\end{figure}

\vspace{-0.07cm}

To parameterize $v_{\boldsymbol{\theta}}(\cdot)$ and $s_{\boldsymbol{\theta}}(\cdot)$ with $\boldsymbol{\theta}$, we propose a two-head \texttt{DiT}-alike architecture that is equipped with the capability of (1) spatiotemporal encoding, and (2) exogenous variable encoding. Fig. \ref{fig.architecture} illustrates \method's architecture.


\method\ extends Diffusion Transformer backbone \citep{peebles2023dit} for spatiotemporal and exogenous modeling. First, each image in the interpolant wave variables, {\em i.e.}, 
$\hat{\mathcal{Y}}_{\tau}$, which is of size $H\times W\times C$, is patchified via a 
$1\times 1$ convolution into 
$N=HW$ patch tokens. 
A linear layer then transforms each token from $C$ channels to $d$ dimensions. All of the $T$ images in $\hat{\mathcal{Y}}_{\tau}$ are then transformed to a tensor of size $(T, N, d)$. Similarly, the atmospheric variables $\mathcal{Z}$ are 
projected to a tensor of size $(T', N', d)$ with a $2\times 2$ convolution, where $N'=H'W'/4$. Additionally, each token is added with a spatial position embedding $\mat{P}_{\text{s}}$ based on sinusoidal function \citep{vaswani2017attention}. The final input wave and atmospheric tokens are $\tilde{\mat{Y}}_{\tau}\in\mathbb{R}^{T\times N\times d}$ and $\tilde{\mat{Z}}\in\mathbb{R}^{T'\times N'\times d}$.

\noindent{\textbf{Conditioning via \texttt{AdaLN-Zero}}}. As illustrated by Fig. \ref{fig.architecture}(a), the wave and atmospheric tokens are fed to the proposed \method\ block, which also takes the flow time $\tau$ in Eq.~\eqref{eq.stochastic} and the physical model's forecasting lead time $T_{l}$ as the condition, enabling the model to be aware of the stochastic process during inference and the difficulty of the forecasting task. Inspired by \texttt{DiT} \citep{peebles2023dit} and \texttt{FiLM} \citep{perez2018film}, we employ adaptive layer norm with zero initialization (\texttt{AdaLN-Zero}) to encode the condition: $\tau$ and $T_{l}$ are mapped through sinusoidal embeddings, concatenated, and passed through a 2-layer \texttt{SiLU} \texttt{MLP} to produce scales ({\em i.e.}, $\boldsymbol{\gamma}$'s in Fig. \ref{fig.architecture}), shifts ({\em i.e.}, $\boldsymbol{\beta}$'s in Fig. \ref{fig.architecture}), and gates ({\em i.e.}, $\boldsymbol{\alpha}$'s in Fig. \ref{fig.architecture}). Particularly, the \texttt{MLP} is initialized to output the zero-vector for all gates $\alpha$'s, ensuring every module in Fig. \ref{fig.architecture}(b) acts as the identity function initially for stable optimization of deep stacks, resembling the skip connections in \texttt{ResNet} \citep{he2016deep}.

\noindent{\textbf{Factorized Attention Modules}}. Each \method\ block applies three attention modules and a feed forward module in sequence:
\begin{equation}
\small
\begin{aligned}
&\tilde{\mat{Y}}^{l,1} = \tilde{\mat{Y}}^{l} + \boldsymbol{\alpha}_{\text{p}}\circ\texttt{SpatialAttention}(\boldsymbol{\gamma}_{\text{p}}\circ\texttt{LayerNorm}(\tilde{\mat{Y}}^{l}) + \boldsymbol{\beta}_{\text{p}})\\
&\tilde{\mat{Y}}^{l,2} = \tilde{\mat{Y}}^{l,1} + \boldsymbol{\alpha}_{\text{c}}\circ\texttt{CrossAttention}(\boldsymbol{\gamma}_{\text{c}}\circ\texttt{LayerNorm}(\tilde{\mat{Y}}^{l,1}) + \boldsymbol{\beta}_{\text{c}}, \tilde{\mat{Z}}) + \mat{P}_{\text{t}}\\
&\tilde{\mat{Y}}^{l,3} = \tilde{\mat{Y}}^{l,2} + \boldsymbol{\alpha}_{\text{t}}\circ\texttt{TemporalAttention}(\boldsymbol{\gamma}_{\text{t}}\circ\texttt{LayerNorm}(\tilde{\mat{Y}}^{l,2}) + \boldsymbol{\beta}_{\text{t}})\\
&\tilde{\mat{Y}}^{l+1} = \tilde{\mat{Y}}^{l,3} + \boldsymbol{\alpha}_{\text{f}}\circ\texttt{FeedForward}(\boldsymbol{\gamma}_{\text{f}}\circ\texttt{LayerNorm}(\tilde{\mat{Y}}^{l,3}) + \boldsymbol{\beta}_{\text{f}})
\end{aligned}
\end{equation}
where $l\in\{0, ..., L-1\}$ denotes the $l$-th \method\ block, and $\tilde{\mat{Y}}^{0} = \tilde{\mat{Y}}_{\tau}$. $\mat{P}_{\text{t}}$ is a temporal position embedding based on sinusoidal function. $\texttt{SpatialAttention}(\cdot)$ applies self-attention to the $N$ patch tokens in $\tilde{\mat{Y}}^{l}\in\mathbb{R}^{T\times N\times d}$ at each time step $t$ ($1\le t\le T$). $\texttt{TemporalAttention}(\cdot)$ applies self-attention to the $T$ patch tokens at the same location across $T$ time steps in parallel. $\texttt{CrossAttention}(\cdot)$ is applied to all patch tokens at each individual time step $t$, which uses scaled/shifted wave patch tokens in $\tilde{\mat{Y}}^{l,1}$ as the Query, and uses atmospheric patch tokens $\tilde{\mat{Z}}$ as the Key and Value. By factorizing spatiotemporal attention, \method\ block avoids $O(N^{2}T^{2})$ cost.

\noindent{\textbf{Two-Head Output}}. The architecture terminates with two independent projection heads for realizing $v_{\boldsymbol{\theta}}(\hat{\mathcal{Y}}_{\tau}, \mathcal{Z}, \tau)$ and $s_{\boldsymbol{\theta}}(\hat{\mathcal{Y}}_{\tau}, \mathcal{Z}, \tau)$ and generating
\begin{equation}
\small
\begin{aligned}
&\mathcal{V} = v_{\boldsymbol{\theta}}(\hat{\mathcal{Y}}_{\tau}, \mathcal{Z}, \tau) = \texttt{Reshape}\big(\texttt{Linear}(\boldsymbol{\gamma}_{\text{v}}\circ\texttt{LayerNorm}(\tilde{\mat{Y}}^{L}) + \boldsymbol{\beta}_{\text{v}})\big)\in\mathbb{R}^{T\times H\times W\times C}\\
&\mathcal{S} = s_{\boldsymbol{\theta}}(\hat{\mathcal{Y}}_{\tau}, \mathcal{Z}, \tau) = \texttt{Reshape}\big(\texttt{Linear}(\boldsymbol{\gamma}_{\text{s}}\circ\texttt{LayerNorm}(\tilde{\mat{Y}}^{L}) + \boldsymbol{\beta}_{\text{s}})\big)\in\mathbb{R}^{T\times H\times W\times C}
\end{aligned}
\end{equation}
which are used in Eq.~\eqref{eq.train_loss} for model training.

\subsection{Training and Inference Algorithms}

The \method\ model learns parameter $\boldsymbol{\theta}$ via Algorithm \ref{alg.train} by minimizing the joint velocity and score loss in Eq~\eqref{eq.train_loss}. In the inference process, we use Euler-Maruyama method \citep{kloeden1992numerical} to derive the forward process for SDE, leading to Algorithm \ref{alg.inference}. For brevity, the full derivation process (involving the Tweedie score) is deferred to Appendix \ref{app.theory}. It is noteworthy that when the SDE noise $\varepsilon_{\text{SDE}}=0$, Algorithm \ref{alg.inference} is equivalent to ODE. Finally, following \citep{lipman2023flow,flowcast_ode,tong2024improving,albergo2025stochastic}, due to the randomness of SDE, 
we generate an ensemble of forecasts to achieve robust probabilistic forecasting.

\begin{algorithm}[!t]
\small
\DontPrintSemicolon
\SetNoFillComment
\KwIn{(1) Training dataset $\mathcal{D}_{\text{tr}}=\{(\mathcal{X}, \mathcal{Y}, \mathcal{Z})\}$, (2) Standard deviation $\sigma$}
\KwOut{Trained model parameter $\boldsymbol{\theta}$}
\BlankLine
Initialize model parameter $\boldsymbol{\theta}$\;
\While{not converged}{
  Sample a batch of data $(\mathcal{X}, \mathcal{Y}, \mathcal{Z}) \sim \mathcal{D}_{\text{tr}}$\;
  Sample $\tau \sim \mathcal{U}(0,1)$ and $\boldsymbol{\epsilon} \sim \mathcal{N}(\mat{0}, \mat{I})$\;
  $\hat{\mathcal{Y}}_{\tau} \leftarrow (1{-}\tau)(\mathcal{X} + \sigma\boldsymbol{\epsilon}) + \tau\mathcal{Y}$\tcp*{Stochastic interpolant in Eq.~\eqref{eq.perturb}}
  $\mathcal{V} \leftarrow v_{\boldsymbol{\theta}}(\hat{\mathcal{Y}}_{\tau}, \mathcal{Z}, \tau)$, $\mathcal{S} \leftarrow s_{\boldsymbol{\theta}}(\hat{\mathcal{Y}}_{\tau}, \mathcal{Z}, \tau)$\tcp*{Forward pass through \method\ (Fig.\ref{fig.architecture}(a))}
  $\ell \leftarrow \|\mathcal{V} - (\mathcal{Y}-\mathcal{X}-\sigma\boldsymbol{\epsilon})\|^{2} + \lambda\|\mathcal{S} - \boldsymbol{\epsilon}\|^{2}$\tcp*{Compute loss in Eq.~\eqref{eq.train_loss}}
  $\boldsymbol{\theta} \leftarrow \mathrm{AdamW}(\ell, \boldsymbol{\theta})$\tcp*{Gradient step}
}
\KwRet $\boldsymbol{\theta}$\;
\caption{\method\ Training Process}\label{alg.train}
\end{algorithm}

\begin{algorithm}[!t]
\small
\DontPrintSemicolon
\SetNoFillComment
\KwIn{(1) Model parameter $\boldsymbol{\theta}$, (2) Sample ($\mathcal{X}$, $\mathcal{Z}$), (3) Standard deviation $\sigma$, (4) Number of SDE steps $S$, (5) Ensemble size $K$, (6) SDE noise level $\varepsilon_{\text{SDE}}$}
\KwOut{Ensemble of forecasts $\{\hat{\mathcal{Y}}^{(k)}\}_{k=1}^{K}$}
\BlankLine
$\Delta\tau \leftarrow 1/S$\tcp*{Set step size}
\For{$k=1, \ldots, K$}{
  Sample $\boldsymbol{\epsilon}^{k} \sim \mathcal{N}(\mat{0}, \mat{I})$\tcp*{Perturbation noise}
  $\hat{\mathcal{Y}}_{\tau} \leftarrow \mathcal{X} + \sigma\boldsymbol{\epsilon}^{(k)}$\tcp*{Initially perturbed wave variables}
  \For{$i = 0, \ldots, S-1$}{
    $\tau_{i} \leftarrow i\cdot\Delta\tau$\tcp*{Set current step}
    $\mathcal{V} \leftarrow v_{\boldsymbol{\theta}}(\hat{\mathcal{Y}}_{\tau}, \mathcal{Z}, \tau_{i})$, $\mathcal{S} \leftarrow s_{\boldsymbol{\theta}}(\hat{\mathcal{Y}}_{\tau}, \mathcal{Z}, \tau_{i})$\tcp*{Forward pass in Fig.\ref{fig.architecture}(a)}
    Tweedie score: $\mathcal{S}_{\text{Tweedie}} \leftarrow -\frac{\mathcal{S}}{(1-\tau_{i})\sigma}$\tcp*{Tweedie score, see Appendix~\ref{app.impl.score}}
    $\hat{\mathcal{Y}}_{\tau} \leftarrow \hat{\mathcal{Y}}_{\tau} + (\mathcal{V} + \varepsilon_{\text{SDE}}\cdot \mathcal{S}_{\text{Tweedie}})\,\Delta\tau$\tcp*{Euler-Maruyama drift step}
    \textbf{If} $i<S-1$: Sample $\boldsymbol{\xi} \sim \mathcal{N}(\mat{0}, \mat{I})$, $\hat{\mathcal{Y}}_{\tau} \leftarrow \hat{\mathcal{Y}}_{\tau} + \sqrt{2\,\varepsilon_{\text{SDE}}\,\Delta\tau}\,\boldsymbol{\xi}$
  }
  $\hat{\mathcal{Y}}^{k} \leftarrow \hat{\mathcal{Y}}_{\tau}$\tcp*{Collect ensemble forecasts}
}
\KwRet $\{\hat{\mathcal{Y}}^{k}\}_{k=1}^{K}$\;
\caption{\method\ Inference Process}\label{alg.inference}
\end{algorithm}

\section{Experiments}\label{sec.exp}

\vspace{-0.1cm}

\textbf{Datasets.} We evaluate \method\ using public real-world oceanic and atmospheric data, including (1) \texttt{ERA5} (ECMWF ReAnalysis, 5th generation) \citep{hersbach2020era5}; (2) \texttt{GFS} (Global Forecast System) \citep{zhou2019toward}; and (3) \texttt{IFS} (Integrated Forecasting System) \citep{richardson2000skill}. \texttt{ERA5} provides reanalysis data about ocean and atmosphere in different regions, which are often considered as accurate and used as ground truth for assessing different forecasting models \citep{hersbach2020era5,sharmar2020validation,zhou2026comparing} (note: \texttt{ERA5} does not perform forecasting). Both \texttt{GFS} and \texttt{IFS} provide wave and atmospheric forecasts by physical models for different regions, but they differ in physical models.

\begin{wraptable}{r}{9.2cm}
\centering
\footnotesize
\setlength{\tabcolsep}{4pt}
\begin{tabular}{|l|l|ccc|c|}\hline
\textbf{Dataset} & \textbf{Region} & $N_{\text{train}}$ & $N_{\text{val}}$ & $N_{\text{test}}$ & \textbf{Dimension} \\ \hline
\multirow{3}{*}{\texttt{GFS}} & Gulf of Mexico & 193,944& 2,904 & 2,904 & $33\times 47$\\
& North Sea & 193,896& 2,904 & 2,904 & $32\times 48$\\
& California Coast & 192,240 & 2,904 & 2,904 & $35\times 41$\\ \hline
\multirow{3}{*}{\texttt{IFS}} & Gulf of Mexico & 29,520 & 17,376 & 17,184 & $33\times 47$\\
& North Sea & 29,520 & 17,376 & 17,184 & $32\times 48$\\
& California Coast & 29,520 & 17,376 & 17,184 & $35\times 41$\\ \hline
\end{tabular}
\end{wraptable}

The collected \texttt{GFS} data is available between 2021 and 2025 
(3hr per frame), which is split into train (2021-2023)/validation (2024)/test (2025) sets. The \texttt{IFS} data is available between 2024 and 2025 (3hr per frame), which is split into train (2024)/validation (2025/1-2025/6)/test (2025/7-2025/12) sets. For both, there are 4 wave variables ({\em e.g.}, wave height, direction) and 11 atmospheric variables ({\em e.g.}, winds, pressure). Samples are generated by segmenting the data into 24-frame windows ({\em i.e.}, 72hr) with stride 2 for \texttt{GFS} and stride 4 for \texttt{IFS}. 
\texttt{GFS}'s validation and tests use stride 24 to compose an evaluation of non-overlapped sample windows. It is noteworthy that each sample may be forecasted by a physical model $T_{l}'$ hrs ahead ($3\le T_{l}'\le 168$), leading to a lead-time $T_{l}\in [3, 240]$ for individual frames in a sample window. We categorize the lead-time into Short-Term ($3\le T_{l}\le 24$), Mid-Term ($24<T_{l}\le 72$), and Long-Term ($72<T_{l}\le 240$) for evaluation in three regions -- Gulf of Mexico, North Sea, and California Coast. The \texttt{ERA5} data covers these periods and are used for model training and evaluation. The resolution of each frame is $0.5^\circ$. The Table above summarizes the datasets, where $N_{\text{train}}$/$N_{\text{val}}$/$N_{\text{test}}$ are the number of frames in 
train/validation/test set. More details about the datasets can be found in Appendix \ref{app.data}.


\noindent{\textbf{The Compared Methods}}. We compare \method\ with the SOTA methods, including the vanilla physical models: (1) \texttt{GFS-WW3} \citep{tolman2009ww3}, (2) \texttt{IFS-WAM} \citep{ecmwf-ifs-wam}; deterministic models: (3) \texttt{U-Net} \citep{ronneberger2015unet}, (4) \texttt{EarthFormer} \citep{gao2022earthformer}, (5) \texttt{EarthFarseer} \citep{wu2024earthfarseer}, (6) \texttt{SimVP} \citep{gao2022simvp}, (7) \texttt{AFNO} \citep{guibas2022afno}, (8) \texttt{TimeXer} \citep{wang2024timexer}; and probabilistic models: (9) \texttt{FlowMatch} \citep{lipman2023flow}, (10) \texttt{DiffCast} \citep{yu2024diffcast}, 
(11) \texttt{FlowCast} \citep{flowcast_ode}. Among them, \texttt{TimeXer} is a recent time series model that can explicitly encode both endogenous variables and exogenous variables. For other baselines, we concatenate the wave and atmospheric variables in their inputs. We trained the baselines to correct physical model' biases as our method does except for \texttt{FlowMatch} which was trained to reconstruct \texttt{ERA5} data without conditioning on physical forecasters. Comparing with it can highlight the importance of bridging physical models to reanalysis.

For our \method, we present its performance using both SDE ($\varepsilon_{\text{SDE}}\neq 0$ in Algorithm \ref{alg.inference}) and ODE ($\varepsilon_{\text{SDE}}=0$) inferences, leading to two variants \method-s ({\em i.e.}, SDE) and \method-o ({\em i.e.}, ODE). The parameter $\sigma$ is grid-searched within $[0, 1]$, $\varepsilon_{\text{SDE}}$ is grid-searched within $(0, 0.1]$ (for a small noise) using validation set. More details about the methods and their parameter settings can be found in Appendix \ref{app.baselines}. The implementation details 
are deferred to Appendix \ref{app.impl_details}.

\noindent{\textbf{Evaluation}}. Following \citep{flowcast_ode}, we evaluate
wave forecasting by widely used metrics, including Root Mean
Square Error (RMSE), Continuous Ranked Probability Score (CRPS)
\citep{gneiting2007proper}, threshold-weighted CRPS (twCRPS)
\citep{fricker2013twcrps}, Critical Success Index (CSI)
\citep{schaefer1990csi}, and Heidke Skill Score (HSS)
\citep{heidke1926hss}. RMSE and CRPS measure forecasting errors, thus a
small value is preferred. CSI and HSS measure hit ratio of correct
predictions, thus a larger value is better. They are particularly useful
for detecting extreme events ({\em e.g.}, storm). twCRPS is the
threshold-weighted CRPS at the 95th percentile that also fits event
detection. Details about the evaluation metrics can be found in
Appendix~\ref{app.evaluation}.

\vspace{-0.1cm}

\subsection{Experimental Results}

\vspace{-0.15cm}

\begin{table*}[!t]
\vspace{-0.3cm}
\centering
\caption{Performance comparison on \texttt{GFS}'s dataset. $\uparrow$ ($\downarrow$) indicates a larger (smaller) value is better. \textbf{\red{Red}} (\underline{\blue{Blue}}) values indicate the best (second-best) values per forecasting term.}\label{tab.exp.gfs}
\vspace{0.1cm}
\tiny
\setlength{\tabcolsep}{1.7pt}{
\begin{tabular}{l|l|ccccc|ccccc|ccccc}
\toprule[1pt]
  &   & \multicolumn{5}{c}{\textbf{Gulf of Mexico}} & \multicolumn{5}{c}{\textbf{North Sea}} & \multicolumn{5}{c}{\textbf{California Coast}} \\
\cmidrule(lr){3-7} \cmidrule(lr){8-12} \cmidrule(lr){13-17}
  & \textbf{Model} & RMSE$_{\downarrow}$ & CRPS$_{\downarrow}$ & CSI$_{\uparrow}$ & HSS$_{\uparrow}$ & twCRPS$_{\downarrow}$ & RMSE$_{\downarrow}$ & CRPS$_{\downarrow}$ & CSI$_{\uparrow}$ & HSS$_{\uparrow}$ & twCRPS$_{\downarrow}$ & RMSE$_{\downarrow}$ & CRPS$_{\downarrow}$ & CSI$_{\uparrow}$ & HSS$_{\uparrow}$ & twCRPS$_{\downarrow}$ \\
\midrule
\multirow{11}{*}{\rotatebox{90}{Short-Term}}
 & \texttt{GFS-WW3} & 0.2351 & 0.1583 & 0.6950 & 0.8096 & 0.0130 & 0.4657 & 0.2851 & 0.7457 & 0.8404 & 0.0434 & 0.2300 & 0.1600 & 0.6510 & 0.7695 & 0.0149 \\
 & \texttt{FlowMatch} & 0.3025 & 0.1404 & 0.3776 & 0.5229 & 0.0261 & 0.7739 & 0.4334 & 0.3358 & 0.4287 & 0.0830 & 0.2966 & 0.1553 & 0.4571 & 0.5881 & 0.0148 \\
& \texttt{U-Net} & 0.1688 & 0.1101 & 0.6986 & 0.8127 & 0.0156 & 0.3563 & 0.2656 & 0.7297 & 0.8299 & 0.0401 & 0.3386 & 0.2504 & 0.5309& 0.6604 & 0.0293 \\
 & \texttt{EarthFormer} & 0.1939 & 0.1173 & 0.6692 & 0.7913 & 0.0187 & 0.3410 & 0.2353 & 0.7092 & 0.8118 & 0.0437 & 0.2314 & 0.1793 & 0.6044 & 0.7298 & 0.0161 \\
 & \texttt{EarthFarseer} & 0.1734 & 0.1096 & 0.6889 & 0.8061 & 0.0164 & 0.2728 & 0.1998 & 0.7775 & 0.8633 & 0.0304 & 0.2128 & 0.1609 & 0.6281 & 0.7515 & 0.0145 \\
 & \texttt{SimVP} & 0.1570 & 0.1086 & 0.7280 & 0.8341 & 0.0123 & 0.3138 & 0.2266 & 0.7494 & 0.8439 & 0.0349 & 0.2361 & 0.1788 & 0.5353 & 0.6444 & 0.0170 \\
 & \texttt{AFNO} & 0.1692 & 0.1104 & 0.7015 & 0.8160 & 0.0141 & 0.2987 & 0.2076 & 0.7475 & 0.8419 & 0.0351 & 0.2074 & 0.1571 & 0.6647 & 0.7838 & 0.0129 \\
 & \texttt{TimeXer} & 0.2008 & 0.1334 & 0.6651 & 0.7886 & 0.0144 & 0.4080 & 0.2631 & 0.7907 & 0.8735 & 0.0288 & 0.2187 & 0.1581 & 0.6731 & 0.7897 & 0.0133 \\
 & \texttt{DiffCast} & 0.1659 & 0.0935 & 0.7074 & 0.8190 & \underline{\blue{0.0103}} & 0.2903 & 0.1617 & 0.7828 & 0.8673 & 0.0240 & 0.2184 & 0.1226 & \textbf{\red{0.7027}} & \textbf{\red{0.8120}} & \textbf{\red{0.0081}} \\
 & \texttt{FlowCast} & 0.2608 & 0.1425 & 0.5689 & 0.7073 & 0.0197 & 0.3337 & 0.1789 & 0.7487 & 0.8436 & 0.0250 & 0.2224 & 0.1212 & 0.6483 & 0.7691 & 0.0106 \\
\cline{2-17}
 & \method-o & \textbf{\red{0.1348}} & \textbf{\red{0.0678}} & \textbf{\red{0.7620}} & \textbf{\red{0.8574}} & \textbf{\red{0.0087}} & \textbf{\red{0.2513}} & \textbf{\red{0.1340}} & \underline{\blue{0.7915}} & \underline{\blue{0.8738}} & \textbf{\red{0.0206}} & \textbf{\red{0.1934}} & \underline{\blue{0.1134}} & \underline{\blue{0.6972}} & \underline{\blue{0.8083}} & 0.0090 \\
 & \method-s & \underline{\blue{0.1359}} & \underline{\blue{0.0681}} & \underline{\blue{0.7591}} & \underline{\blue{0.8554}} & \textbf{\red{0.0087}} & \underline{\blue{0.2526}} & \underline{\blue{0.1350}} & \textbf{\red{0.7937}} & \textbf{\red{0.8755}} & \underline{\blue{0.0207}} & \underline{\blue{0.1942}} & \textbf{\red{0.1070}} & 0.6900 & 0.8026 & \underline{\blue{0.0084}} \\
\midrule
\multirow{11}{*}{\rotatebox{90}{Mid-Term}}
 & \texttt{GFS-WW3} & 0.2554 & 0.1783 & 0.5815 & 0.7218 & 0.0121 & 0.5459 & 0.3516 & 0.6551 & 0.7743 & 0.0465 & 0.2625 & 0.1789 & 0.6488 & 0.7778 & 0.0142 \\
  & \texttt{FlowMatch} & 0.2651 & 0.1382 & 0.2995 & 0.4234 & 0.0154 & 0.7462 & 0.4195 & 0.2631 & 0.3542 & 0.0617 & 0.3131 & 0.1624 & 0.3536 & 0.4735 & 0.0242 \\
& \texttt{U-Net} & 0.1626 & 0.1133 & 0.6255 & 0.7576 & 0.0105 & 0.4313 & 0.3117 & 0.6212 & 0.7462 & 0.0406 & 0.3455 & 0.2547 & 0.5299& 0.6779 & 0.0213 \\
& \texttt{EarthFormer} & 0.1706 & 0.1175 & 0.5701 & 0.7104 & 0.0119 & 0.4190 & 0.2921 & 0.5788 & 0.7038 & 0.0430 & 0.2414 & 0.1819 & 0.5985 & 0.7384 & 0.0129 \\
 & \texttt{EarthFarseer} & 0.1669 & 0.1156 & 0.6122 & 0.7481 & 0.0109 & 0.3943 & 0.2739 & 0.6161 & 0.7394 & 0.0390 & 0.2364 & 0.1736 & 0.5695 & 0.7088 & 0.0145 \\
 & \texttt{SimVP} & 0.1738 & 0.1219 & 0.6139 & 0.7490 & 0.0107 & 0.4113 & 0.2885 & 0.6067 & 0.7310 & 0.0407 & 0.2502 & 0.1890 & 0.5898 & 0.7312 & 0.0129 \\
 & \texttt{AFNO} & 0.1713 & 0.1183 & 0.5817 & 0.7220 & 0.0111 & 0.3886 & 0.2694 & 0.6317 & 0.7539 & 0.0365 & 0.2272 & 0.1708 & 0.6651 & 0.7908 & 0.0110 \\
 & \texttt{TimeXer} & 0.2098 & 0.1447 & 0.5487 & 0.6943 & 0.0120 & 0.4966 & 0.3357 & 0.6342 & 0.7574 & 0.0369 & 0.2451 & 0.1763 & 0.6436 & 0.7745 & 0.0113 \\
 & \texttt{DiffCast} & 0.1890 & 0.1068 & 0.5762 & 0.7181 & 0.0100 & 0.3900 & 0.2098 & \textbf{\red{0.6814}} & \textbf{\red{0.7953}} & 0.0269 & 0.2447 & 0.1359 & \underline{\blue{0.6777}} & \underline{\blue{0.7999}} & \underline{\blue{0.0084}} \\
 & \texttt{FlowCast} & 0.2367 & 0.1362 & 0.5545 & 0.6966 & 0.0134 & 0.4040 & \underline{\blue{0.2090}} & 0.6507 & 0.7708 & 0.0268 & 0.2416 & 0.1278 & 0.6290 & 0.7617 & 0.0103 \\
\cline{2-17}
 & \method-o & \textbf{\red{0.1551}} & \underline{\blue{0.0797}} & \textbf{\red{0.6610}} & \textbf{\red{0.7860}} & \underline{\blue{0.0076}} & \textbf{\red{0.3531}} & \textbf{\red{0.1784}} & \underline{\blue{0.6702}} & \underline{\blue{0.7857}} & \textbf{\red{0.0266}} & \textbf{\red{0.2146}} & \underline{\blue{0.1251}} & \textbf{\red{0.6788}} & \textbf{\red{0.8011}} & 0.0089 \\
 & \method-s & \underline{\blue{0.1564}} & \textbf{\red{0.0795}} & \underline{\blue{0.6593}} & \underline{\blue{0.7847}} & \textbf{\red{0.0073}} & \underline{\blue{0.3538}} & \textbf{\red{0.1784}} & 0.6696 & 0.7852 & \underline{\blue{0.0267}} & \underline{\blue{0.2171}} & \textbf{\red{0.1170}} & 0.6712 & 0.7956 & \textbf{\red{0.0082}} \\
\midrule
\multirow{11}{*}{\rotatebox{90}{Long-Term}}
 & \texttt{GFS-WW3} & 0.3617 & 0.2434 & 0.4250 & 0.5681 & 0.0280 & 0.7411 & 0.4903 & \underline{\blue{0.3295}} & \underline{\blue{0.4702}} & 0.0312 & 0.4293 & 0.2692 & 0.4243 & 0.5662 & 0.0353 \\
  & \texttt{FlowMatch} & 0.3402 & 0.1718 & 0.2792 & 0.3896 & 0.0240 & 0.7522 & 0.4291 & 0.1340 & 0.2090 & 0.0206 & 0.4155 & 0.2089 & 0.2793 & 0.3909 & 0.0263 \\
& \texttt{U-Net} & 0.2928 & 0.1817 & \textbf{\red{0.4502}} & \textbf{\red{0.5906}} & 0.0273 & 0.6224 & 0.4229 & 0.2939 & 0.4233 & 0.0236 & 0.4713 & 0.3261 & 0.4030& 0.5441 & 0.0380 \\ 
 & \texttt{EarthFormer} & 0.2963 & 0.1828 & 0.4124 & 0.5508 & 0.0265 & \textbf{\red{0.5735}} & 0.3843 & 0.2589 & 0.3679 & 0.0209 & \underline{\blue{0.3788}} & 0.2646 & 0.3989 & 0.5273 & \underline{\blue{0.0243}} \\
 & \texttt{EarthFarseer} & 0.2932 & 0.1836 & 0.4211 & 0.5630 & 0.0266 & 0.5789 & 0.3849 & 0.2619 & 0.3738 & 0.0216 & 0.3862 & 0.2630 & 0.4150 & 0.5573 & 0.0266 \\
 & \texttt{SimVP} & 0.3021 & 0.1893 & 0.4209 & 0.5611 & 0.0270 & 0.5837 & 0.3925 & 0.2680 & 0.3872 & 0.0208 & 0.3831 & 0.2655 & 0.4108 & 0.5467 & 0.0260 \\
 & \texttt{AFNO} & 0.2968 & 0.1846 & 0.4178 & 0.5610 & 0.0257 & 0.5818 & 0.3807 & 0.2924 & 0.4213 & 0.0217 & 0.3827 & 0.2605 & \textbf{\red{0.4480}} & \underline{\blue{0.5917}} & 0.0263 \\
 & \texttt{TimeXer} & 0.3201 & 0.2077 & 0.3922 & 0.5354 & 0.0262 & 0.6650 & 0.4527 & 0.3131 & 0.4528 & 0.0214 & 0.4010 & 0.2683 & 0.3887 & 0.5215 & 0.0279 \\
 & \texttt{DiffCast} & 0.3131 & 0.1712 & 0.4216 & 0.5647 & 0.0256 & 0.5942 & 0.3081 & \textbf{\red{0.3368}} & \textbf{\red{0.4786}} & 0.0195 & 0.4012 & 0.2160 & 0.4401 & 0.5831 & 0.0247 \\
 & \texttt{FlowCast} & 0.3452 & 0.1753 & 0.4266 & 0.5656 & 0.0264 & 0.6132 & 0.3075 & 0.3129 & 0.4477 & 0.0188 & 0.3969 & 0.2084 & \textbf{\red{0.4480}} & \textbf{\red{0.5929}} & 0.0244 \\
\cline{2-17}
 & \method-o & \textbf{\red{0.2916}} & \underline{\blue{0.1460}} & \underline{\blue{0.4433}} & \underline{\blue{0.5864}} & \underline{\blue{0.0227}} & \underline{\blue{0.5743}} & \underline{\blue{0.2919}} & 0.2915 & 0.4209 & \underline{\blue{0.0180}} & \textbf{\red{0.3755}} & \underline{\blue{0.2071}} & \underline{\blue{0.4445}} & 0.5880 & 0.0245 \\
 & \method-s & \underline{\blue{0.2924}} & \textbf{\red{0.1388}} & 0.4421 & 0.5851 & \textbf{\red{0.0214}} & 0.5753 & \textbf{\red{0.2867}} & 0.2949 & 0.4247 & \textbf{\red{0.0175}} & \underline{\blue{0.3788}} & \textbf{\red{0.1881}} & 0.4406 & 0.5838 & \textbf{\red{0.0226}} \\
\midrule
\multirow{11}{*}{\rotatebox{90}{Overall}}
 & \texttt{GFS-WW3} & 0.2993 & 0.2013 & 0.5244 & 0.6701 & 0.0184 & 0.6402 & 0.4119 & 0.5713 & 0.7105 & 0.0386 & 0.3334 & 0.2114 & 0.5446 & 0.6870 & 0.0225 \\
 & \texttt{FlowMatch} & 0.3072 & 0.1551 & 0.2918 & 0.4128 & 0.0201 & 0.7593 & 0.4304 & 0.2288 & 0.3238 & 0.0411 & 0.3869 & 0.1933 & 0.3269 & 0.4486 & 0.0230 \\
 & \texttt{U-Net} & 0.2228 & 0.1394 & 0.5512 & 0.6923 & 0.0176 & 0.5260 & 0.3609 & 0.5500 & 0.6926 & 0.0323 & 0.3976 & 0.2814 & 0.4744 & 0.6205 & 0.0288 \\
 & \texttt{EarthFormer} & 0.2301 & 0.1429 & 0.5087 & 0.6531 & 0.0184 & 0.4927 & 0.3309 & 0.5166 & 0.6583 & 0.0322 & 0.3005 & 0.2133 & 0.5116 & 0.6544 & 0.0178 \\
 & \texttt{EarthFarseer} & 0.2253 & 0.1414 & 0.5316 & 0.6767 & 0.0176 & \underline{\blue{0.4834}} & 0.3204 & 0.5508 & 0.6923 & 0.0296 & 0.3006 & 0.2065 & 0.5097 & 0.6515 & 0.0193 \\
 & \texttt{SimVP} & 0.2308 & 0.1466 & 0.5393 & 0.6833 & 0.0172 & 0.4943 & 0.3327 & 0.5432 & 0.6856 & 0.0303 & 0.3067 & 0.2172 & 0.5046 & 0.6479 & 0.0186 \\
 & \texttt{AFNO} & 0.2282 & 0.1432 & 0.5217 & 0.6680 & 0.0171 & 0.4839 & 0.3169 & 0.5574 & 0.6986 & 0.0290 & 0.2949 & 0.2036 & 0.5641 & 0.7050 & 0.0172 \\
 & \texttt{TimeXer} & 0.2575 & 0.1679 & 0.4929 & 0.6423 & 0.0178 & 0.5764 & 0.3851 & 0.5720 & 0.7128 & 0.0285 & 0.3120 & 0.2096 & 0.5274 & 0.6718 & 0.0181 \\
 & \texttt{DiffCast} & 0.2430 & 0.1304 & 0.5187 & 0.6654 & 0.0161 & 0.4915 & 0.2525 & \textbf{\red{0.6062}} & \textbf{\red{0.7410}} & 0.0230 & 0.3116 & 0.1650 & \underline{\blue{0.5687}} & \underline{\blue{0.7090}} & \underline{\blue{0.0147}} \\
 & \texttt{FlowCast} & 0.2863 & 0.1521 & 0.4980 & 0.6426 & 0.0191 & 0.5098 & 0.2534 & 0.5730 & 0.7126 & 0.0227 & 0.3085 & 0.1578 & 0.5470 & 0.6897 & 0.0158 \\
\cline{2-17}
 & \method-o & \textbf{\red{0.2169}} & \underline{\blue{0.1041}} & \textbf{\red{0.5727}} & \textbf{\red{0.7132}} & \underline{\blue{0.0136}} & \textbf{\red{0.4654}} & \underline{\blue{0.2286}} & 0.5915 & 0.7291 & \underline{\blue{0.0217}} & \textbf{\red{0.2853}} & \underline{\blue{0.1553}} & \textbf{\red{0.5696}} & \textbf{\red{0.7098}} & 0.0150 \\
 & \method-s & \underline{\blue{0.2185}} & \textbf{\red{0.1027}} & \underline{\blue{0.5697}} & \underline{\blue{0.7106}} & \textbf{\red{0.0131}} & \textbf{\red{0.4654}} & \textbf{\red{0.2274}} & \underline{\blue{0.5917}} & \underline{\blue{0.7292}} & \textbf{\red{0.0214}} & \underline{\blue{0.2872}} & \textbf{\red{0.1436}} & 0.5650 & 0.7058 & \textbf{\red{0.0139}} \\
\bottomrule[1pt]
\end{tabular}
}
\vspace{-0.1cm}
\end{table*}

\begin{figure}[!t]
\centering
\includegraphics[width=\linewidth]{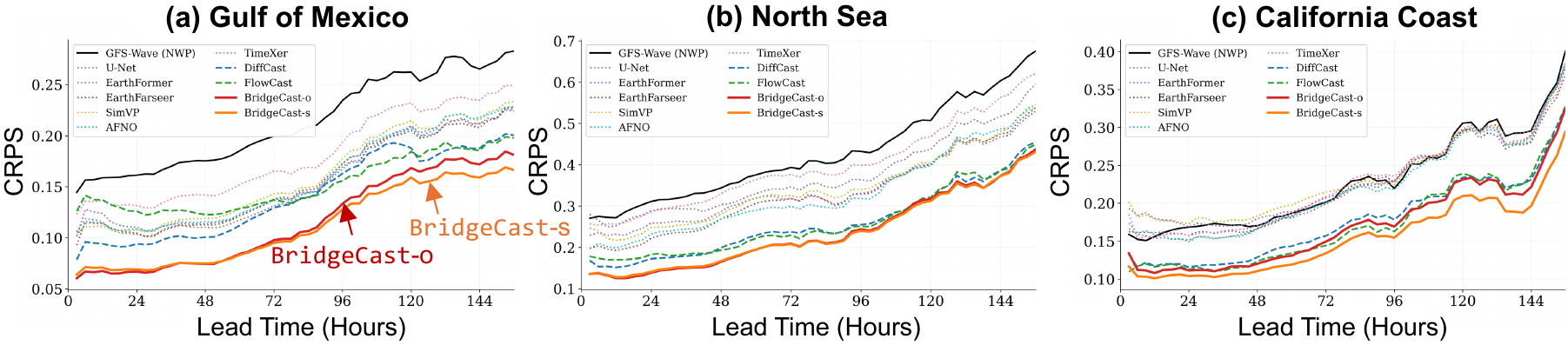}
\caption{CRPS comparison {\em w.r.t.} different lead times across different regions.}\label{fig.crps}
\vspace{-0.2cm}
\end{figure}

Table \ref{tab.exp.gfs} summarizes the bias correction performance of the compared methods on \texttt{GFS}'s 
wave heights across three regions and three forecasting terms, along with an overall performance without differentiating forecasting terms. For brevity, the results on other variables are deferred to Appendix \ref{app.results.mwpd_gfs}, and the results on \texttt{IFS}'s datasets are deferred to Appendix \ref{app.results.ifs}, which show similar insights. In Table \ref{tab.exp.gfs}, several key insights emerge: (1) All models perform better for shorter lead-times; (2) Compared with \texttt{GFS-WW3}, the improvements of other models suggests applying an AI-based model for correcting the physical model's forecasting bias is useful; (3) the inferior performance of \texttt{FlowMatch} highlights the importance to bridge physical forecasts to \texttt{ERA5}; (4) Deterministic models are generally inferior to probabilistic models in terms of CRPS and twCRPS, which assess their ability in modeling uncertainty; (5) Notably, \method, which is theoretically designed for bias correction and architecturally devised for spatiotemporal modeling with exogenous variables, outperforms the baselines in most cases, confirming its effectiveness; and (6) While \method-o performs best in most metrics, \method-s achieves the best CRPS and twCRPS, which is further confirmed by its consistent superiority in CRPS in Fig. \ref{fig.crps}, demonstrating \method\ can be easily switched by setting $\varepsilon_{\text{SDE}}=0$ (\method-o) and $\varepsilon_{\text{SDE}}\neq 0$ (\method-s) at inference time (without retraining) based on the requirement for uncertainty-aware forecasting.

\vspace{-0.1cm}

\subsection{Ablation Analysis}

\vspace{-0.1cm}

We validate the design of \method\ through ablation studies using \texttt{GFS}'s dataset in Gulf of Mexico. Table \ref{tab.exp.ablation} summarizes the analysis of \method-o (\method-s uses the same trained parameter $\boldsymbol{\theta}$ thus is omitted for brevity): (a) removes exogenous atmospheric variables $\mathcal{Z}$, along with the associated cross-attention module in Fig. \ref{fig.architecture}; 
(b) removes the spatial attention in Fig. \ref{fig.architecture}; (c) removes temporal attention ; (d) removes the lead-time $T_{l}$ from the condition input in Fig. \ref{fig.architecture}(a); (e) removes the denoiser head $s_{\boldsymbol{\theta}}(\cdot)$ in Fig. \ref{fig.architecture}(a); (f) removes the perturbation $\boldsymbol{\epsilon}$ on $\mathcal{X}$ in Eq.~\eqref{eq.perturb} by setting $\sigma=0$; and (g) switches the flow matching model by diffusion model DDIM \citep{song2021denoising}.

\begin{table*}[!t]
\vspace{-0.3cm}
\centering
\caption{Ablation Analysis of \method-o on \texttt{GFS}'s dataset in Gulf of Mexico. $\uparrow$ ($\downarrow$) indicates a larger (smaller) value is better. ``Improvement'' of each ablation is relative to \method-o.}\label{tab.exp.ablation}
\vspace{0.1cm}
\tiny
\setlength{\tabcolsep}{1.5pt}{
\begin{tabular}{l|cccc|cccc|cccc}
\toprule[1pt]
& \multicolumn{4}{c}{\textbf{Short-Term}} & \multicolumn{4}{c}{\textbf{Mid-Term}} & \multicolumn{4}{c}{\textbf{Long-Term}} \\
\cmidrule(lr){2-5} \cmidrule(lr){6-9} \cmidrule(lr){10-13}
\textbf{Model} & RMSE$_{\downarrow}$ & CRPS$_{\downarrow}$ & CSI$_{\uparrow}$ & HSS$_{\uparrow}$ & RMSE$_{\downarrow}$ & CRPS$_{\downarrow}$ & CSI$_{\uparrow}$ & HSS$_{\uparrow}$ & RMSE$_{\downarrow}$ & CRPS$_{\downarrow}$ & CSI$_{\uparrow}$ & HSS$_{\uparrow}$ \\
\midrule
\method-o & 0.1348 & 0.0678 & 0.7620 & 0.8574 & 0.1551 & 0.0797 & 0.6610 & 0.7860 & 0.2916 & 0.1460 & 0.4433 & 0.5864 \\ \hline
(a) - Exogenous variables $\mathcal{Z}$ & +0.1403 & 0.0689 & 0.7514 & 0.8523 & 0.1558 & 0.0819 & 0.6602 & 0.7880 & 0.2908 & 0.1466 & 0.4522 & 0.6013 \\
~~~~~~Improvement & \red{-4.08\%} & \red{-1.62\%} & \red{-1.39\%} & \red{-0.59\%} & \red{-0.44\%} & \red{-2.76\%} & \red{-0.13\%} & +0.25\% & +0.26\% & \red{-0.41\%} & +2.03\% & +2.54\% \\ \hline
(b) - Spatial attention module & 0.1517 & 0.0732 & 0.7451 & 0.8479 & 0.1603 & 0.0774 & 0.6459 & 0.7768 & 0.3019 & 0.1372 & 0.4204 & 0.5667 \\
~~~~~~Improvement & \red{-12.52\%} & \red{-7.93\%} & \red{-2.22\%} & \red{-1.11\%} & \red{-3.31\%} & +2.89\% & \red{-2.29\%} & \red{-1.17\%} & \red{-3.56\%} & +6.04\% & \red{-5.17\%} & \red{-3.35\%} \\ \hline
(c) - Temporal attention & 0.1433 & 0.0773 & 0.7391 & 0.8435 & 0.1673 & 0.0918 & 0.6531 & 0.7822 & 0.3104 & 0.1641 & 0.4606 & 0.6084 \\
~~~~~~Improvement & \red{-6.30\%} & \red{-14.01\%} & \red{-3.01\%} & \red{-1.62\%} & \red{-7.87\%} & \red{-15.18\%} & \red{-1.20\%} & \red{-0.48\%} & \red{-6.45\%} & \red{-12.40\%} & +3.90\% & +3.75\% \\ \hline
(d) - Lead-time $T_{l}$ in condition & 0.1517 & 0.0793 & 0.7358 & 0.8417 & 0.1703 & 0.0926 & 0.6276 & 0.7631 & 0.2970 & 0.1568 & 0.4390 & 0.5876 \\
~~~~~~Improvement & \red{-12.51\%} & \red{-17.04\%} & \red{-3.44\%} & \red{-1.83\%} & \red{-9.77\%} & \red{-16.22\%} & \red{-5.06\%} & \red{-2.92\%} & \red{-1.87\%} & \red{-7.38\%} & \red{-0.98\%} & +0.20\% \\ \hline
(e) - Denoiser head $s_{\boldsymbol{\theta}}(\cdot)$ & 0.1360 & 0.0695 & 0.7568 & 0.8557 & 0.1574 & 0.0830 & 0.6618 & 0.7892 & 0.2906 & 0.1497 & 0.4464 & 0.5935 \\
~~~~~~Improvement & \red{-0.83\%} & \red{-2.52\%} & \red{-0.69\%} & \red{-0.20\%} & \red{-1.46\%} & \red{-4.11\%} & +0.12\% & +0.41\% & +0.33\% & \red{-2.57\%} & +0.72\% & +1.22\% \\ \hline
(f) - Perturbation ($\sigma=0$ in Eq.~\eqref{eq.perturb}) & 0.1340 & 0.0911 & 0.7525 & 0.8530 & 0.1660 & 0.1120 & 0.6351 & 0.7685 & 0.3191 & 0.1934 & 0.4218 & 0.5686 \\
~~~~~~Improvement & +0.61\% & \red{-34.47\%} & \red{-1.25\%} & \red{-0.51\%} & \red{-6.99\%} & \red{-40.54\%} & \red{-3.93\%} & \red{-2.23\%} & \red{-9.43\%} & \red{-32.47\%} & \red{-4.84\%} & \red{-3.03\%} \\ \hline
(g) Flow matching$\rightarrow$DDIM & 0.6307 & 0.3463 & 0.2559 & 0.3715 & 0.7780 & 0.4486 & 0.1629 & 0.2364 & 0.8693 & 0.4976 & 0.1461 & 0.2010 \\
~~~~~~Improvement & \red{-367.71\%} & \red{-411.04\%} & \red{-66.43\%} & \red{-56.66\%} & \red{-401.42\%} & \red{-462.96\%} & \red{-75.36\%} & \red{-69.93\%} & \red{-198.13\%} & \red{-240.83\%} & \red{-67.04\%} & \red{-65.73\%} \\
\bottomrule[1pt]
\end{tabular}
}
\vspace{-0.1cm}
\end{table*}

Table \ref{tab.exp.ablation} reveals key insights into \method's design. (a) suggests the usefulness of atmospheric variables, though slight as they have already been incorporated by physical models. (b)(c) indicate the importance of spatiotemporal modeling, while (d) shows conditioning on $T_{l}$ enables \method\ to be aware of lead-time. The denoiser $s_{\boldsymbol{\theta}}(\cdot)$ is beneficial in most cases as a regularizer of the model. (f) and (g) indicate perturbation and flow matching are critical to the performance of \method.

\subsection{Performance Analysis}

\vspace{-0.1cm}

\begin{figure}[!t]
\centering
\includegraphics[width=\linewidth]{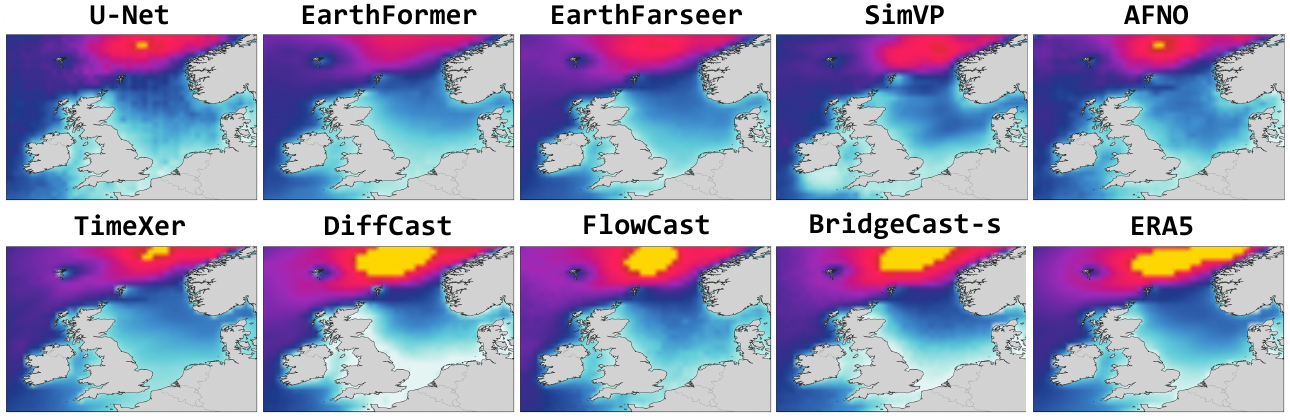}
\caption{Qualitative comparison of \method\ and other baselines on \texttt{GFS}'s data in North Sea region (lead time 24h) when there was a storm event.}\label{fig.vis}
\vspace{-0.2cm}
\end{figure}

\noindent{\textbf{Qualitative Results}}. Fig. \ref{fig.vis} qualitatively compares \method-s (\method-o is similar thus omitted for brevity) with the baselines on \texttt{GFS}'s data in North Sea region (lead time 24h) when there was a storm event. \texttt{FlowMatch} is excluded due to its inferior performance. \method-s produces realistic forecasts that align better with the \texttt{ERA5} reanalysis data. Other models mostly under-forecast the wave during storm. The 
\texttt{DiffCast}'s forecast, though appears reasonable, actually shifted when comparing with \texttt{ERA5}, leading to a higher CRPS (0.39) than \method-s (0.22).

\begin{wraptable}{r}{7cm}
\centering
\caption{Inference Time of \method.}\label{tab:inference_cost_grid}
\vspace{0.1cm}
\small
\setlength{\tabcolsep}{4pt}{
\begin{tabular}{l|l|ccc}
\toprule
\textbf{Data} & \textbf{Model} & \textbf{Gulf (s)} & \textbf{NS (s)} & \textbf{Cali (s)} \\
\midrule
\multirow{2}{*}{\texttt{GFS}} & \method-o & 0.714 & 0.681 & 0.636 \\
                     & \method-s & 0.743 & 0.710 & 0.665 \\
\midrule
\multirow{2}{*}{\texttt{IFS}} & \method-o & 0.951 & 0.914 & 0.840 \\
                     & \method-s & 0.980 & 0.942 & 0.874 \\
\bottomrule
\end{tabular}
}
\end{wraptable}


\noindent{\textbf{Inference Time}}. Table \ref{tab:inference_cost_grid} summarizes the averaged per-sample (a 24-frames window) inference time of \method\ across three regions for both \texttt{GFS} and \texttt{IFS} data. 
\method-s is slightly slower than \method-o. However, both models' inference time are relatively marginal {\em w.r.t.} the update time of physical models \texttt{GFS-WW3} ({\em i.e.}, 6h) and \texttt{IFS-WAM} ({\em i.e.}, 12h), 
confirming the practical feasibility of \method\ as a plug-and-play module for operational systems for ocean wave forecasting.

\noindent{\textbf{Parameter Analysis}}. We have evaluated the impacts of parameters $\sigma$ and $\varepsilon_{\text{SDE}}$. For brevity, these results are deferred to Appendix \ref{app.results.sweep}, which demonstrate \method\ is not sensitive to them.

\vspace{-0.1cm}

\section{Conclusion}

\vspace{-0.1cm}

In this work, we proposed \method\ within a physics-AI 
framework for ocean wave forecasting. 
\method\ learns to bridge 
physics-based forecasts toward that of reanalysis data through a 
CFM process. The proposed Transformer-based architecture supports spatiotemporal modeling, exogenous variable encoding, and flexible ODE/SDE inference. Extensive experiments on \texttt{GFS}, \texttt{IFS}, and \texttt{ERA5} datasets demonstrate that \method\ outperforms SOTA deterministic and probabilistic baselines across different lead times and different regions. These results highlight the potential of \method\ for improving operational geoscientific forecasting systems as a flexible plug-and-play module. Additionally, we defer our discussion of limitation and future directions to Appendix \ref{app.limitations}.

\newpage

\section*{AI Use Statement}

In this work, we have not used generative AI tools for the tasks with required disclosure, and the rest of the required disclosure tasks are not applicable to this work. Additionally, we used generative AI tools for editing the paper to improve readability. 
We have reviewed all AI-assisted work. We take responsibility for the final content of this work, including text, claims or artifacts produced with the aid of generative AI.

\section*{Reproducibility Statement}

In this paper, $\S$\ref{sec:method} describes the framework of \method. $\S$\ref{sec.exp} describes the implementation and configuration of \method. We are committed to releasing the code used in our experiments to facilitate the reproducibility of this work.

\bibliographystyle{iclr2027_conference}
\bibliography{ref}

\newpage
\appendix
\section*{Appendix}\label{app}

\section{Limitations}\label{app.limitations}

\noindent{\textbf{Two Inference Modes}}. The proposed \method\ has two inference methods, including SDE-based inference ($\varepsilon_{\text{SDE}}\neq 0$), leading to \method-s, and ODE-based inference ($\varepsilon_{\text{SDE}}=0$), leading to \method-o. \method-s is suitable for probabilistic forecasting but is less efficient than \method-o, which also supports probabilistic forecasting through random initial perturbations. While the proposed model support the two approaches based on application requirements, it cannot provide the advantage of both methods simultaneously. A future work is to investigate the trade-off between efficiency and uncertainty modeling for a balanced development.

\noindent{\textbf{Inadequate Modeling of Look-Back Windows}}. The proposed \method\ correct biases of physical models instantly. The current design has not yet incorporated the historical wave and atmospheric data as look-back windows for forecasting-style bias correction. Due to the temporal dependency such as periodicity in the data, modeling look-back windows together with physical model outputs and exogenous variables may provide further improvement, which remains open for further exploration.




\section{Theoretical Development}\label{app.theory}

This appendix derives the joint training objective in 
Eq.~\eqref{eq.train_loss} and the inference algorithm in 
Algorithm~\ref{alg.inference} from the Stochastic Interpolant (SI) 
framework of \citep{albergo2025si}, which unifies flow matching 
\citep{lipman2023flow} and score-based diffusion 
\citep{song2021score,ho2020}. We instantiate \method{} as an SI 
with data-dependent coupling ($\S$\ref{app.impl.si}), derive 
the velocity loss ($\S$\ref{app.impl.si}), recover the score from a 
denoiser via Tweedie's identity ($\S$\ref{app.impl.score}), assemble the 
joint objective ($\S$\ref{app.impl.score}), and derive the SDE sampler 
($\S$\ref{app.impl.sde}).

\subsection{\method{} as a Stochastic Interpolant}\label{app.impl.si}

We instantiate \method{} as a stochastic interpolant
\citep[Definition~1]{albergo2025si}: a stochastic process $\{x_\tau\}_{\tau\in[0,1]}$
of the form $x_\tau = I(\tau, x_0, x_1) + \gamma(\tau)z$ with
$(x_0, x_1)\sim\nu$, $z\sim\mathcal{N}(0, I)$, boundary conditions
$I(0)=x_0$, $I(1)=x_1$, and $\gamma(0)=\gamma(1)=0$.

\paragraph{\method{} instantiation.}
We choose
\begin{equation}\label{eq.app.bc_choice}
I(\tau, \tilde{x}_0, x_1) = (1-\tau)\tilde{x}_0 + \tau x_1, \quad
\gamma \equiv 0, \quad
\tilde{x}_0 = \mathcal{X} + \sigma\boldsymbol{\epsilon}, \quad
x_1 = \mathcal{Y},
\end{equation}
which yields the \method{} interpolant
\begin{equation}\label{eq.app.bc_interp}
\hat{\mathcal{Y}}_\tau = (1-\tau)(\mathcal{X} + \sigma\boldsymbol{\epsilon}) + \tau\mathcal{Y},
\end{equation}
under the data dependent coupling \citep{albergo2023datadep}
\begin{equation}\label{eq.app.coupling}
\nu(d\tilde{x}_0, dx_1) = \int \mathcal{N}(\tilde{x}_0; \mathcal{X}, \sigma^2 I)\,\pi(d\mathcal{X}, d\mathcal{Y}),
\end{equation}
where $\pi$ is the empirical joint of paired forecast--reanalysis fields.
The induced marginals are $\rho_0 = \pi_{\mathcal{X}} * \mathcal{N}(0, \sigma^2 I)$
and $\rho_1 = \pi_{\mathcal{Y}}$. The Gaussian smoothed source $\rho_0$
satisfies the regularity condition required for SI admissibility
\citep[Remark~11]{albergo2025si}, so all results we invoke from
\citep{albergo2025si} apply. In particular, conditional on
$(\mathcal{X}, \mathcal{Y})$,
\begin{equation}\label{eq.app.condgauss}
\hat{\mathcal{Y}}_\tau \mid \mathcal{X}, \mathcal{Y} \sim
\mathcal{N}(\boldsymbol{\mu}_\tau, \sigma_\tau^2 I), \quad
\boldsymbol{\mu}_\tau = (1-\tau)\mathcal{X} + \tau\mathcal{Y}, \quad
\sigma_\tau = (1-\tau)\sigma.
\end{equation}

\paragraph{Velocity loss.}
The SI velocity field is defined as the conditional expectation of the
pathwise time derivative,
\begin{equation}\label{eq.app.v_def}
v(\tau, x) \;:=\; \mathbb{E}\!\left[\partial_\tau \hat{\mathcal{Y}}_\tau \;\big|\; \hat{\mathcal{Y}}_\tau = x\right].
\end{equation}
Differentiating \eqref{eq.app.bc_interp} pathwise gives
\begin{equation}\label{eq.app.vstar}
\partial_\tau \hat{\mathcal{Y}}_\tau \;=\; \mathcal{Y} - \mathcal{X} - \sigma\boldsymbol{\epsilon}.
\end{equation}
We learn $v$ via $L^2$ regression of a parameterized field $v_\theta$
against this pathwise target:
\begin{equation}\label{eq.app.vobj}
\mathcal{L}_v[v_\theta] \;=\; \mathbb{E}\,\big\|v_\theta(\hat{\mathcal{Y}}_\tau, \mathcal{X}, \mathcal{Z}, \tau)
- (\mathcal{Y} - \mathcal{X} - \sigma\boldsymbol{\epsilon})\big\|^2.
\end{equation}
The unique minimizer of \eqref{eq.app.vobj} over measurable functions of
$\hat{\mathcal{Y}}_\tau$ is the conditional expectation
\eqref{eq.app.v_def}; this is the standard $L^2$-projection property
\citep[Theorem~7]{albergo2025si}.


\subsection{Score via Tweedie's Identity, Denoiser Loss, and Joint Objective}\label{app.impl.score}

\paragraph{Tweedie's identity} \citep{robbins1956empirical,efron2011tweedie}. Let $X = \boldsymbol{\mu} + \kappa\boldsymbol{\epsilon}$ with
$\boldsymbol{\mu}\sim p(\boldsymbol{\mu})$,
$\boldsymbol{\epsilon}\sim\mathcal{N}(\mat{0},\mat{I})\perp\boldsymbol{\mu}$,
and $\kappa>0$. Then
\begin{equation}\label{eq.app.tweedie}
\mathbb{E}[\boldsymbol{\epsilon} \mid X = x] = -\kappa\,\nabla_x \log p_X(x).
\end{equation}
\textit{Proof.} Differentiating
$p_X(x) = \int p(\boldsymbol{\mu})\,\mathcal{N}(x; \boldsymbol{\mu}, \kappa^2\mat{I})\,d\boldsymbol{\mu}$
under the integral and dividing by $p_X(x)$:
\begin{equation*}
\nabla_x \log p_X(x) = \frac{\mathbb{E}[\boldsymbol{\mu}\mid x] - x}{\kappa^2}.
\end{equation*}
Since $\boldsymbol{\epsilon} = (X - \boldsymbol{\mu})/\kappa$, $\mathbb{E}[\boldsymbol{\epsilon}\mid x] = -\kappa\,\nabla_x\log p_X(x)$.

\paragraph{Application to \method{}.}
The conditional Gaussian \eqref{eq.app.condgauss} matches the Tweedie
hypothesis with $\kappa = \sigma_\tau$. Applying \eqref{eq.app.tweedie}
to the marginal $\rho_\tau$ yields
\begin{equation}\label{eq.app.score_id}
\nabla_x \log \rho_\tau(x) = -\frac{1}{\sigma_\tau}\,
\mathbb{E}[\boldsymbol{\epsilon} \mid \hat{\mathcal{Y}}_\tau = x],
\quad \tau\in[0,1),
\end{equation}
which matches \citep[Eq.~2.18]{albergo2025si} under
$\gamma(\tau)\leftrightarrow\sigma_\tau$.

\paragraph{Denoiser loss.}
Following noise-prediction diffusion \citep{ho2020}, we train
$\eta_\theta$ to regress $\boldsymbol{\epsilon}$:
\begin{equation}\label{eq.app.eta_obj}
\mathcal{L}_\eta[\eta_\theta] = \mathbb{E}\,
\big\|\eta_\theta(\hat{\mathcal{Y}}_\tau,  \mathcal{Z}, \tau)
- \boldsymbol{\epsilon}\big\|^2.
\end{equation}
The $L^2$ minimizer is
$\eta_\theta^\star = \mathbb{E}[\boldsymbol{\epsilon}\mid\hat{\mathcal{Y}}_\tau]$.
Combined with \eqref{eq.app.score_id}, the score is recovered at
inference as
\begin{equation}\label{eq.app.score_recover}
\nabla_x \log \rho_\tau(x) = -\frac{
\eta_\theta(\hat{\mathcal{Y}}_\tau, \mathcal{Z}, \tau)}{\sigma_\tau}.
\end{equation}
This is denoising score matching at noise level $\sigma_\tau$
\citep{vincent2011}. The denoiser parameterization is preferred over
direct score regression since the latter's target diverges as
$\sigma_\tau^{-1}$ as $\tau\to 1$ \citep[Remark~9]{albergo2025si},
whereas $\boldsymbol{\epsilon}\sim\mathcal{N}(\mat{0},\mat{I})$has a finite, time-independent second moment.

\paragraph{Joint objective.}
Combining \eqref{eq.app.vobj} (with antithetic sampling) and
\eqref{eq.app.eta_obj}:
\begin{equation}\label{eq.app.joint}
\mathcal{L}(\boldsymbol{\theta}) = \mathbb{E}\!\left[
\big\|v_\theta - (\mathcal{Y} - \mathcal{X}-\sigma\boldsymbol{\epsilon})\big\|^2
+ \big\|\eta_\theta - \boldsymbol{\epsilon}\big\|^2\right],
\end{equation}
where $v_\theta, \eta_\theta$ share inputs
$(\hat{\mathcal{Y}}_\tau 
, \mathcal{Z}, \tau)$ and the
expectation is over
$(\mathcal{X}, \mathcal{Y}, \mathcal{Z})\sim\mathcal{D}_{\text{tr}}$,
$\tau\sim\mathcal{U}([0,1])$, and
$\boldsymbol{\epsilon}\sim\mathcal{N}(\mat{0},\mat{I})$. This is
Eq.~\eqref{eq.train_loss}.

\subsection{Inference SDE}\label{app.impl.sde}
By \citep[Corollary~10]{albergo2025si} (cf. \citealp{song2021score} 
for the score-based diffusion analog), for any $\varepsilon\geq 0$, 
the SDE become,
\begin{equation}\label{eq.app.sde_family}
d\hat{\mathcal{Y}}_\tau = [v + \varepsilon\, s]\,d\tau + \sqrt{2\varepsilon}\,dW_\tau
\end{equation}
preserves the marginals $\{\rho_\tau\}$ shared with the probability-flow 
ODE $d\hat{\mathcal{Y}}_\tau = v\,d\tau$. Substituting the score 
recovery \eqref{eq.app.score_recover}:
\begin{equation}\label{eq.app.sde_bc}
d\hat{\mathcal{Y}}_\tau = \!\left[v_\theta - \frac{\varepsilon_{\text{SDE}}}{\sigma_\tau}\,\eta_\theta\right]\!d\tau + \sqrt{2\varepsilon_{\text{SDE}}}\,dW_\tau.
\end{equation}
Euler--Maruyama discretization with $\Delta\tau = 1/S$, step index 
$i\in\{0,\ldots,S-1\}$, $\tau_i = i\Delta\tau$:
\begin{equation}\label{eq.app.em_step}
\hat{\mathcal{Y}}_{\tau_{i+1}} = \hat{\mathcal{Y}}_{\tau_i} + \!\left[v_\theta - \frac{\varepsilon_{\text{SDE}}}{\sigma_{\tau_i}}\eta_\theta\right]\!\Delta\tau + \sqrt{2\varepsilon_{\text{SDE}}\Delta\tau}\,\boldsymbol{\xi}_i,
\end{equation}
with $\boldsymbol{\xi}_i\sim\mathcal{N}(\mat{0},\mat{I})$ drawn fresh 
per step. Per member initial conditions 
$\hat{\mathcal{Y}}_0 = \mathcal{X} + \sigma\boldsymbol{\epsilon}^{(k)}$ 
provide ensemble spread; $\varepsilon_{\text{SDE}}=0$ recovers the 
probability-flow ODE ($\eta_\theta$ is unused in this limit). A 
numerical floor $\sigma_\tau \geq 10^{-3}$ prevents division by zero 
as $\tau\to 1$ \citep[Algorithm~5]{albergo2025si}. This yields 
Algorithm~\ref{alg.inference}.

\section{Experiments}

\subsection{Evaluation Metrics}\label{app.evaluation}

\paragraph{Root Mean Square Error (RMSE).}
Computed on the ensemble mean $\bar{\mathcal{Y}}$ for stochastic methods
and on the single prediction for deterministic methods:
\begin{equation}\label{eq.app.rmse}
\text{RMSE} = \sqrt{\frac{1}{|\Omega|}\sum_{i\in\Omega}\bigl(\bar{\mathcal{Y}}_i - \mathcal{Y}_i\bigr)^2},
\qquad
\bar{\mathcal{Y}}_{i} = \frac{1}{K}\sum_{k=1}^K \hat{\mathcal{Y}}_{i}^{(k)}
\end{equation}
where $\Omega$ denotes the set of ocean pixels. Lower is better.

\paragraph{Continuous Ranked Probability Score (CRPS).}
Let 
$F$ be a predictive cumulative distribution function
(CDF), 
and 
$y$ be an 
observed target value. 
CRPS is defined as
\begin{equation}\label{eq.app.crps}
\mathrm{CRPS}(F, y)
=
\int_{-\infty}^{\infty}
\left[
F(x)-\mathds{1}(x\geq y)
\right]^2\,dx,
\end{equation}
where $F(x)$ is the CDF of the predicted probability distribution and $\mathds{1}(x\geq y)$ is an indicator function equal to 1 when $x\geq y$ and 0 otherwise. The indicator function represents the CDF of the observed value, which is a step function. CRPS measures the integrated squared difference between the predicted CDF and this step function. Thus, a lower CRPS indicates a better forecast that assigns probability toward the observed target value. 

As can be seen, CRPS evaluates the entire predictive distribution, rather than just a point prediction, making it useful for probabilistic forecasting tasks.

For a deterministic forecast, {\em i.e.}, $K=1$ where $K$ is the ensemble
size, that predicts a single value $\hat{y}$, CRPS reduces to the absolute error $|\hat{y} - y|$ \citep{gneiting2007proper}. Averaging the absolute errors over multiple samples yields the mean absolute error (MAE). Thus, CRPS generalizes MAE.



\paragraph{Threshold-Weighted CRPS (twCRPS).}
A weighted variant of CRPS that emphasizes performance on extreme
events \citep{fricker2013twcrps}. Let $q_{95}$ denote the train-set
$95$th percentile of SWH. The twCRPS uses a chaining function
$v(x) = \max(x, q_{95})$ to penalize errors only above the threshold:
\begin{equation}\label{eq.app.twcrps}
\text{twCRPS} = \mathbb{E}_{\hat{y}}\bigl|v(\hat{y}) - v(y)\bigr|
- \tfrac{1}{2}\,\mathbb{E}_{\hat{y},\hat{y}'}\bigl|v(\hat{y}) - v(\hat{y}')\bigr|,
\end{equation}

\paragraph{Critical Success Index (CSI).}
A categorical skill score for binary event detection
\citep{schaefer1990csi}. We threshold the ensemble mean and ground
truth at the train-set $95$th percentile to obtain binary masks, then
compute the contingency table (true positives $\text{TP}$, false
positives $\text{FP}$, false negatives $\text{FN}$):
\begin{equation}\label{eq.app.csi}
\text{CSI} = \frac{\text{TP}}{\text{TP} + \text{FP} + \text{FN}}.
\end{equation}
CSI ranges in $[0,1]$, rewards correct event hits, and penalizes both
missed and false alarms. It is widely used in nowcasting evaluation
\citep{flowcast_ode}. 

\paragraph{Heidke Skill Score (HSS).}
A categorical skill score that accounts for chance agreement
\citep{heidke1926hss}. Using the same contingency table:
\begin{equation}\label{eq.app.hss}
\text{HSS} = \frac{2(\text{TP}\cdot\text{TN} - \text{FP}\cdot\text{FN})}
                  {(\text{TP}+\text{FN})(\text{FN}+\text{TN}) + (\text{TP}+\text{FP})(\text{FP}+\text{TN})}.
\end{equation}
HSS is bounded above by $1$; values above $0$ indicate skill exceeding
a random forecast. 

\paragraph{Mean Wave Period and Direction CRPS.}
For the auxiliary wave variables we use channel-specific CRPS variants.
\textbf{CRPS-MWP} (s) applies \eqref{eq.app.crps} to mean wave period.
\textbf{CRPS-MWD} ($^\circ$) is the mean angular distance between the
ensemble-mean direction $\bar{\theta}$ (recovered from the $(\sin,\cos)$
ensemble means) and the target $\theta^*$:
\begin{equation}\label{eq.app.mwd}
\text{CRPS-MWD} = \min\bigl(\lvert\bar{\theta} - \theta^*\rvert,\ 
                            360^\circ - \lvert\bar{\theta} - \theta^*\rvert\bigr).
\end{equation}
Both are reported per channel.

\subsection{Baselines}\label{app.baselines}

\paragraph{Baselines.}
We compare \method{} against nine representative baselines spanning
deterministic spatiotemporal forecasters and probabilistic
generative models, all trained on the same paired
$(\mathcal{X}, \mathcal{Z}, \mathcal{Y})$ tuples and evaluated under
the same protocol.

The deterministic baselines cover four major architectural families
relevant to gridded geoscience forecasting.
\texttt{U-Net}~\citep{ronneberger2015unet} is the canonical
encoder--decoder convolutional baseline, configured with base channel
width 64 (GFS) or 32 (IFS) to match training-set size.
\texttt{SimVP}~\citep{gao2022simvp} uses depthwise-separable
spatial--temporal convolutional blocks for parameter-efficient video
prediction.
\texttt{AFNO}~\citep{guibas2022afno} replaces self-attention with
adaptive Fourier neural operators and is the standard
spectral-mixing baseline used in modern weather emulators
({\em e.g.}, \texttt{FourCastNet}).
\texttt{EarthFormer}~\citep{gao2022earthformer} introduces cuboid
attention along factorized spatial and temporal axes; we use the
$[4,4,4]$ cuboid configuration tuned for regional grids.
\texttt{EarthFarseer}~\citep{wu2024earthfarseer} augments a
SimVP-style backbone with a Fourier-Transformer (FoTF) module to
capture long-range dependencies; it is our most parameter-heavy
deterministic baseline ($\sim$25.6\,M).
\texttt{TimeXer}~\citep{wang2024timexer} is a transformer that
explicitly separates exogenous (atmospheric) and endogenous (wave)
channels through cross-attention with a residual prediction head ---
the closest deterministic analog to our two-head design.
 
The probabilistic baselines fall into two families.
\texttt{DiffCast}~\citep{yu2024diffcast} is a denoising diffusion
model trained from Gaussian noise to the target distribution with
1000 training steps and 250-step DDIM sampling at inference; we use
its standard \texttt{U-Net} backbone with channel multipliers $(1,2,4)$.
\texttt{FlowCast} is a Conditional Flow Matching
\citep{lipman2023flow} variant we construct on top of an
EarthFormer backbone, sampled with a 10-step Euler ODE solver; it
shares the bridge-style training objective with \method{} but uses
a Gaussian source rather than a forecast-anchored source.
 
\method{} is comparable in parameter count to \texttt{FlowCast} and lighter
than \texttt{DiffCast} and \texttt{EarthFarseer}. 
This places \method's gains squarely on the modeling side rather than
on raw scale.

\subsection{Data Details}\label{app.data}

\subsubsection{Forecast Archives}\label{app.data.archives}

\paragraph{\texttt{GFS-WW3} (primary).}
We use the operational NCEP GFS-WW3 (WAVEWATCH~III)
forecast~\citep{tolman2009ww3}, taking all four daily
initialization cycles (00, 06, 12, 18\,UTC) and extracting four
wave channels: significant wave height (SWH), mean wave period
(MWP), and mean wave direction encoded as
$(\sin\theta_d,\cos\theta_d)$ to avoid the $0/2\pi$ discontinuity.
\texttt{GFS-WW3} is distributed at $0.5^\circ$ on the regional crops we
use, matching the \texttt{ERA5} wave grid which is produced only at
$0.5^\circ$~\citep{hersbach2020era5}.
The regional \texttt{GFS-WW3} archive begins in March 2021 and bounds the
GFS training-set size.

\paragraph{\texttt{IFS-WAM} (cross-system).}
We additionally use ECMWF IFS-WAM forecasts~\citep{ecmwf-ifs-wam} from the public
ECMWF open-data archive at 00 and 12\,UTC, lead times
$0,3,\ldots,240$\,h.
Native \texttt{IFS-WAM} resolution is $0.25^\circ$; we stride-2 subsample
to $0.5^\circ$ on the same regional bounding boxes as \texttt{GFS-WW3} so
the input grid is bit-identical to the GFS training grid.
The IFS open-data archive begins 2024-03-01, which limits the IFS
training budget to roughly one seventh of the GFS budget --- a
deliberate stress test of the corrector in the small-data regime.

\paragraph{Atmospheric exogenous $\mathcal{Z}$.}
For each forecast system, atmospheric fields are taken from the
same forecast cycle and lead time as the wave forecast and
regridded to the wave grid before training, so $\mathcal{Z}$ and
$\mathcal{X}$ share an identical $H{\times}W$
spatial extent and no spatial resampling occurs inside the
network.
$\mathcal{Z}$ contains $C'=11$ channels: raw 10\,m wind components
$(u_{10},v_{10})$ and mean sea-level pressure (msl), plus eight
deterministic derivations chosen to make wind-forcing physics
visible to the network --- 10\,m wind speed, wind direction
$(\sin,\cos)$, $u_{10}^{2}$, the per-frame finite-difference time
derivatives of $u_{10}$, $v_{10}$, and msl, and the wind-forcing
magnitude $u_{10}^{2}+v_{10}^{2}$.
The derived channels are mathematically redundant given the raw
fields but accelerate convergence on the small training budgets
used here, following the inductive-bias argument made by \texttt{CorrDiff}
\citep{mardani2025corrdiff} for its derived input fields.
$\mathcal{Z}$ enters \method\ as conditioning only and is never
part of the supervised target.

\paragraph{Target $\mathcal{Y}$.}
The supervised target is the corresponding \texttt{ERA5} wave field~\citep{hersbach2020era5}
at the same valid times and the same $0.5^\circ$ grid, with the same four
channels.
\texttt{ERA5} wave variables are assimilated against altimeter and buoy
observations and are the standard reference for wave-model
verification.

\subsubsection{Regions and Sample Construction}\label{app.data.regions}

\paragraph{Study regions.}
We evaluate on three coastal regions spanning different wave
climates: the \textbf{Gulf of Mexico} ($15.0$--$31.0^\circ$\,N,
$98.0$--$75.0^\circ$\,W; grid $33{\times}47$;
$\approx 0.85$ ocean), a fetch-limited basin dominated by
tropical cyclones; the \textbf{North Sea}
($48.0$--$63.5^\circ$\,N, $4.0^\circ$\,W--$19.5^\circ$\,E;
grid $32{\times}48$; $\approx 0.78$ ocean), an extratropical-storm
regime with the highest baseline GFS-WW3 errors of the three;
and the \textbf{California coast} ($30.0$--$47.0^\circ$\,N,
$135.0$--$115.0^\circ$\,W; grid $35{\times}41$;
$\approx 0.81$ ocean), a long-fetch swell regime in which the raw
forecast is already comparatively strong and residual-correction
gains are hardest to obtain.
Land cells are excluded from every loss and every metric, for
\method\ and for every baseline.

\paragraph{Sample geometry.}
Each sample is a trajectory of $T=24$ frames at
$\Delta t=3$\,h, starting at forecast lead $\ell\in[3,168]$\,h.
Frame index $s\in\{0,\ldots,T-1\}$ has valid forecast lead
$T_{l,s}=\ell+s\Delta t$.
The trajectory length is fixed across samples, while the
window-start lead varies. Individual frames are assigned to
the lead-time bands defined below, whose upper boundary is
240\,h.

\paragraph{Lead-time conditioning: one model, all leads.}
\method\ trains a single model jointly across forecast lead times,
conditioning on a per-frame valid-lead encoding
$\tilde\ell_s=(\ell+s\Delta t)/\ell_{\max}$,
where $\ell_{\max}=168$\,h denotes the maximum window-start lead.
This encoding is supplied to \texttt{AdaLN-Zero} alongside the
interpolant time $\tau$.

\paragraph{Lead-time sampling rationale.}
Operational use of marine forecasts concentrates at short leads
(next-day decisions for shipping, port operations, offshore work,
and fisheries dominate query traffic).
A uniform-in-$\ell$ training distribution would over-represent
long leads relative to where forecast skill matters in deployment.
We therefore sample $\ell\sim\log\,\mathcal{U}[3,168]\,$h on the
3\,h grid, which (i) concentrates training mass in proportion to
operational utility and (ii) acts as importance-sampling
reweighting of the joint loss across leads --- the same role that
non-uniform timestep sampling plays in modern diffusion training,
where it has been shown to reduce gradient variance and accelerate
convergence~\citep{ho2020,song2021score}.

\texttt{GFS-WW3} (NOMADS, $0.5^\circ$, four daily cycles since
March 2021) materializes the $\ell$-sampler at export time:
$\ell\sim\log\,\mathcal{U}[3,168]\,$h is drawn independently per
cycle, yielding a variable count of trajectories per cycle (mean
$\approx 2.3$, range 1--10).
The variability reflects the per-cycle stochastic draw and the
constraint that cycles late in the archive period drop trajectories
at long $\ell$ where the \texttt{ERA5} target window extends past the cache
end.
Validation and test use a 72\,h non-overlapping initialization
stride for independent samples.

\texttt{IFS-WAM} (ECMWF Open Data via AWS public replica bucket,
two daily cycles since 2024-03-01) is much smaller: the official
ECMWF Open Data Portal retains only the most recent $\sim$12
forecast runs, and the full MARS historical archive requires a
commercial Service Agreement we do not hold.
This caps the IFS training budget at roughly $1/7$ the GFS budget.
Given the small archive, we adopt deterministic per-cycle export
with exactly two $\ell$ draws per (00/12\,UTC) initialization from
the 3\,h grid $\{3, 6, \ldots, 168\}\,$h.
Validation and test retain every available initialization;
trajectories therefore overlap in valid time (median 3\,h spacing),
and test-set metrics aggregate over unique initialization times to
avoid double-counting overlapping windows.

\paragraph{Lead-time bands.}
We report results in three operationally meaningful bands chosen
\emph{a priori}: \textbf{Short-Term} (3--24\,h),
\textbf{Mid-Term} (27--72\,h), and \textbf{Long-Term}
(75--240\,h).

\subsubsection{Splits and Sample Counts}\label{app.data.splits}

Splits are strictly chronological.
For \texttt{GFS-WW3}: training 2021-03--2023-12, validation 2024, test
2025, yielding $\sim$8{,}080 training trajectories per region
(Gulf 8{,}081; NS 8{,}079; California 8{,}010) and 121
non-overlapping validation and 121 test trajectories per region.
For \texttt{IFS-WAM}: training 2024-03--2024-12, validation
2025-01--2025-06, test 2025-07--2025-12, yielding
1{,}230 training, 724 validation, and 716 test trajectories
per region.
Normalization statistics are computed from the training split only
and frozen for validation and test.

\subsection{Additional Results}\label{app.results}

\subsubsection{\texttt{IFS-WAM} to \texttt{ERA5} Benchmark}\label{app.results.ifs}

Table~\ref{tab.exp.ifs} demonstrates that \method{} generalizes beyond
\texttt{GFS}: when adapted to the \texttt{IFS-WAM} forecast system,
the model systematically corrects forecast errors across all three
regions and three lead bands and the overall evaluation. Both \method-o and \method-s remain
top 2 across nearly every (region, lead band) cell.

\begin{table*}[!t]
\centering
\caption{Performance comparison on \texttt{IFS}'s dataset. $\uparrow$ ($\downarrow$) indicates a larger (smaller) value is better. \textbf{\red{Red}} (\underline{\blue{Blue}}) values indicate the best (second-best) values per forecasting term.}\label{tab.exp.ifs}
\vspace{0.1cm}
\tiny
\setlength{\tabcolsep}{1.7pt}{
\begin{tabular}{l|l|ccccc|ccccc|ccccc}
\toprule[1pt]
  &   & \multicolumn{5}{c}{\textbf{Gulf of Mexico}} & \multicolumn{5}{c}{\textbf{North Sea}} & \multicolumn{5}{c}{{\textbf{California Coast}}} \\
\cmidrule(lr){3-7} \cmidrule(lr){8-12} \cmidrule(lr){13-17}
  & \textbf{Model} & RMSE$_{\downarrow}$ & CRPS$_{\downarrow}$ & CSI$_{\uparrow}$ & HSS$_{\uparrow}$ & twCRPS$_{\downarrow}$ & RMSE$_{\downarrow}$ & CRPS$_{\downarrow}$ & CSI$_{\uparrow}$ & HSS$_{\uparrow}$ & twCRPS$_{\downarrow}$ & RMSE$_{\downarrow}$ & CRPS$_{\downarrow}$ & CSI$_{\uparrow}$ & HSS$_{\uparrow}$ & twCRPS$_{\downarrow}$ \\
\midrule
\multirow{11}{*}{\rotatebox{90}{Short-Term}}
 & \texttt{IFS-WAM} & 0.2110 & 0.1397 & 0.6829 & 0.8015 & 0.0133 & 0.4328 & 0.2771 & 0.6630 & 0.7855 & 0.0202 & 0.2342 & 0.1676 & 0.6385 & 0.7659 & 0.0181 \\
 
 & \texttt{U-Net} & 0.1564 & 0.0936 & 0.6889 & 0.8076 & 0.0123 & 0.2687 & 0.1944 & 0.5982 & 0.7229 & 0.0178 & 0.1908 & 0.1302 & 0.6596 & 0.7836 & 0.0130 \\
 & \texttt{EarthFormer} & 0.1738 & 0.1067 & 0.6049 & 0.7380 & 0.0152 & 0.2940 & 0.2111 & 0.5712 & 0.6992 & 0.0213 & 0.2030 & 0.1412 & 0.6088 & 0.7357 & 0.0146 \\
 & \texttt{EarthFarseer} & 0.1372 & 0.0874 & 0.6808 & 0.8014 & 0.0118 & 0.2559 & 0.1794 & 0.6006 & 0.7274 & 0.0176 & 0.1820 & 0.1274 & 0.6578 & 0.7799 & 0.0124 \\
 & \texttt{SimVP} & 0.1758 & 0.1117 & 0.6398 & 0.7698 & 0.0136 & 0.3256 & 0.2387 & 0.6331 & 0.7615 & 0.0171 & 0.2122 & 0.1560 & 0.6696 & 0.7897 & 0.0130 \\
 & \texttt{AFNO} & 0.1540 & 0.0980 & 0.7150 & 0.8266 & 0.0113 & 0.2928 & 0.2119 & 0.6274 & 0.7533 & 0.0164 & 0.1821 & 0.1206 & 0.6883 & 0.8036 & 0.0116 \\
 & \texttt{TimeXer} & 0.1590 & 0.1002 & 0.7651 & 0.8611 & 0.0083 & 0.3529 & 0.2067 & 0.7373 & 0.8407 & 0.0115 & 0.1733 & 0.1115 & \textbf{\red{0.7746}} & \textbf{\red{0.8672}} & 0.0080 \\
 & \texttt{DiffCast} & 0.1936 & 0.1127 & 0.7329 & 0.8385 & 0.0069 & 0.4789 & 0.2790 & 0.6678 & 0.7908 & 0.0115 & 0.2326 & 0.1448 & 0.7141 & 0.8251 & 0.0079 \\
 & \texttt{FlowCast} & 0.1770 & 0.0922 & 0.6870 & 0.8058 & 0.0096 & 0.3907 & 0.2170 & 0.5323 & 0.6723 & 0.0206 & 0.1382 & 0.0724 & 0.7487 & 0.8476 & 0.0065 \\
\cline{2-17}
 & \method-o & \textbf{\red{0.1072}} & \textbf{\red{0.0527}} & \textbf{\red{0.7808}} & \textbf{\red{0.8719}} & \underline{\blue{0.0061}} & \textbf{\red{0.1829}} & \textbf{\red{0.0957}} & \textbf{\red{0.7619}} & \textbf{\red{0.8578}} & \underline{\blue{0.0079}} & \textbf{\red{0.1296}} & \underline{\blue{0.0671}} & 0.7738 & 0.8660 & \underline{\blue{0.0063}} \\
 & \method-s & \underline{\blue{0.1107}} & \underline{\blue{0.0619}} & \underline{\blue{0.7785}} & \underline{\blue{0.8704}} & \textbf{\red{0.0057}} & \underline{\blue{0.1834}} & \underline{\blue{0.0961}} & \underline{\blue{0.7594}} & \underline{\blue{0.8560}} & \textbf{\red{0.0078}} & \underline{\blue{0.1298}} & \textbf{\red{0.0665}} & \underline{\blue{0.7744}} & \underline{\blue{0.8664}} & \textbf{\red{0.0062}} \\
\midrule
\multirow{11}{*}{\rotatebox{90}{Mid-Term}}
 & \texttt{IFS-WAM} & 0.2298 & 0.1515 & 0.6431 & 0.7727 & 0.0143 & 0.4705 & 0.3054 & 0.6164 & 0.7474 & 0.0288 & 0.2501 & 0.1708 & 0.6090 & 0.7428 & 0.0172 \\
 
 & \texttt{U-Net} & 0.1721 & 0.1049 & 0.6271 & 0.7602 & 0.0129 & 0.3136 & \underline{\blue{0.2174}} & 0.5900 & 0.7192 & 0.0223 & 0.2054 & 0.1359 & 0.5862 & 0.7209 & 0.0137 \\
 & \texttt{EarthFormer} & 0.1932 & 0.1166 & 0.5036 & 0.6380 & 0.0165 & 0.3405 & 0.2349 & 0.5531 & 0.6855 & 0.0252 & 0.2267 & 0.1502 & 0.5103 & 0.6456 & 0.0155 \\
 & \texttt{EarthFarseer} & 0.1794 & 0.1100 & 0.5840 & 0.7240 & 0.0140 & 0.3286 & 0.2215 & 0.5439 & 0.6780 & 0.0246 & 0.2200 & 0.1461 & 0.5241 & 0.6610 & 0.0142 \\
 & \texttt{SimVP} & 0.1972 & 0.1255 & 0.5375 & 0.6816 & 0.0152 & 0.3620 & 0.2589 & 0.6127 & 0.7444 & 0.0225 & 0.2254 & 0.1609 & 0.5869 & 0.7208 & 0.0137 \\
 & \texttt{AFNO} & 0.1751 & 0.1109 & 0.6340 & 0.7657 & 0.0125 & 0.3296 & 0.2328 & 0.6298 & 0.7585 & 0.0202 & 0.2036 & 0.1335 & 0.5918 & 0.7232 & 0.0126 \\
 & \texttt{TimeXer} & 0.1880 & 0.1195 & 0.6734 & 0.7970 & 0.0112 & 0.4064 & 0.2533 & 0.6623 & 0.7851 & 0.0191 & 0.2062 & 0.1319 & 0.6381 & 0.7656 & 0.0112 \\
 & \texttt{DiffCast} & 0.2491 & 0.1393 & 0.6569 & 0.7848 & \underline{\blue{0.0092}} & 0.4627 & 0.2668 & 0.6335 & 0.7625 & 0.0167 & 0.2460 & 0.1487 & 0.6316 & 0.7621 & 0.0096 \\
 & \texttt{FlowCast} & 0.1951 & 0.1010 & 0.6428 & 0.7729 & 0.0098 & 0.4177 & 0.2256 & 0.5370 & 0.6771 & 0.0227 & 0.1688 & 0.0827 & 0.6479 & 0.7695 & 0.0092 \\
\cline{2-17}
 & \method-o & \textbf{\red{0.1465}} & \textbf{\red{0.0699}} & \textbf{\red{0.6875}} & \textbf{\red{0.8071}} & \underline{\blue{0.0092}} & \textbf{\red{0.2627}} & \textbf{\red{0.1331}} & \textbf{\red{0.6956}} & \textbf{\red{0.8102}} & \underline{\blue{0.0141}} & \textbf{\red{0.1627}} & \underline{\blue{0.0825}} & \textbf{\red{0.6635}} & \textbf{\red{0.7844}} & \underline{\blue{0.0091}} \\
 & \method-s & \underline{\blue{0.1487}} & \underline{\blue{0.0740}} & \underline{\blue{0.6862}} & \underline{\blue{0.8061}} & \textbf{\red{0.0085}} & \underline{\blue{0.2649}} & \textbf{\red{0.1331}} & \underline{\blue{0.6949}} & \underline{\blue{0.8097}} & \textbf{\red{0.0135}} & \underline{\blue{0.1638}} & \textbf{\red{0.0809}} & \underline{\blue{0.6623}} & \underline{\blue{0.7834}} & \textbf{\red{0.0088}} \\
\midrule
\multirow{11}{*}{\rotatebox{90}{Long-Term}}
 & \texttt{IFS-WAM} & 0.3200 & 0.2059 & 0.5205 & 0.6718 & 0.0185 & 0.7497 & 0.4870 & 0.4374 & 0.5838 & 0.0461 & 0.4004 & 0.2548 & 0.4431 & 0.5860 & 0.0242 \\
 
 & \texttt{U-Net} & 0.2703 & 0.1573 & 0.4640 & 0.6191 & 0.0172 & 0.6267 & 0.3940 & 0.4155 & 0.5596 & 0.0352 & 0.3532 & 0.2147 & 0.4424 & 0.5881 & 0.0190 \\
 & \texttt{EarthFormer} & 0.2581 & 0.1537 & 0.3730 & 0.5033 & 0.0195 & 0.5891 & 0.3869 & 0.3661 & 0.4956 & 0.0369 & 0.3197 & 0.2057 & 0.3989 & 0.5367 & 0.0191 \\
 & \texttt{EarthFarseer} & 0.2604 & 0.1563 & 0.4227 & 0.5738 & 0.0184 & 0.6132 & 0.3974 & 0.3687 & 0.5071 & 0.0383 & 0.3217 & 0.2062 & 0.3833 & 0.5133 & 0.0192 \\
 & \texttt{SimVP} & 0.2636 & 0.1645 & 0.4095 & 0.5575 & 0.0188 & 0.5905 & 0.4026 & 0.4316 & 0.5789 & 0.0361 & 0.3201 & 0.2153 & 0.4485 & 0.5919 & 0.0190 \\
 & \texttt{AFNO} & 0.2770 & 0.1650 & 0.4861 & 0.6407 & 0.0166 & 0.6472 & 0.4082 & 0.4319 & 0.5799 & 0.0341 & 0.3623 & 0.2168 & 0.4302 & 0.5763 & 0.0181 \\
 & \texttt{TimeXer} & 0.2605 & 0.1621 & 0.5138 & 0.6664 & 0.0165 & 0.6315 & 0.4152 & 0.4427 & 0.5905 & 0.0348 & 0.3134 & 0.1976 & 0.4655 & 0.6085 & 0.0180 \\
 & \texttt{DiffCast} & 0.3126 & 0.1644 & 0.5092 & 0.6637 & 0.0141 & 0.7034 & 0.3765 & 0.4362 & 0.5852 & 0.0317 & 0.3552 & 0.1899 & 0.4511 & 0.5980 & 0.0160 \\
 & \texttt{FlowCast} & 0.2720 & 0.1310 & 0.5024 & 0.6556 & \underline{\blue{0.0140}} & 0.6636 & 0.3325 & 0.3924 & 0.5314 & 0.0331 & \textbf{\red{0.2849}} & \textbf{\red{0.1377}} & 0.4759 & 0.6177 & \textbf{\red{0.0153}} \\
\cline{2-17}
 & \method-o & \textbf{\red{0.2296}} & \underline{\blue{0.1097}} & \textbf{\red{0.5285}} & \textbf{\red{0.6786}} & 0.0144 & \textbf{\red{0.5619}} & \underline{\blue{0.2945}} & \underline{\blue{0.4626}} & \underline{\blue{0.6098}} & \underline{\blue{0.0315}} & \underline{\blue{0.2954}} & 0.1524 & \textbf{\red{0.4786}} & \textbf{\red{0.6211}} & 0.0175 \\
 & \method-s & \underline{\blue{0.2307}} & \textbf{\red{0.1061}} & \underline{\blue{0.5278}} & \underline{\blue{0.6779}} & \textbf{\red{0.0135}} & \underline{\blue{0.5689}} & \textbf{\red{0.2816}} & \textbf{\red{0.4629}} & \textbf{\red{0.6099}} & \textbf{\red{0.0296}} & 0.3003 & \underline{\blue{0.1463}} & \underline{\blue{0.4774}} & \underline{\blue{0.6200}} & \underline{\blue{0.0159}} \\
\midrule
\multirow{11}{*}{\rotatebox{90}{Overall}}
 & \texttt{IFS-WAM} & 0.2639 & 0.1698 & 0.6055 & 0.7433 & 0.0157 & 0.5839 & 0.3681 & 0.5526 & 0.6934 & 0.0342 & 0.3108 & 0.2007 & 0.5537 & 0.6942 & 0.0199 \\
 
 & \texttt{U-Net} & 0.2113 & 0.1225 & 0.5774 & 0.7203 & 0.0144 & 0.4493 & 0.2788 & 0.5237 & 0.6620 & 0.0266 & 0.2665 & 0.1636 & 0.5456 & 0.6866 & 0.0155 \\
 & \texttt{EarthFormer} & 0.2169 & 0.1289 & 0.4705 & 0.6082 & 0.0174 & 0.4434 & 0.2875 & 0.4846 & 0.6212 & 0.0291 & 0.2620 & 0.1694 & 0.4836 & 0.6220 & 0.0167 \\
 & \texttt{EarthFarseer} & 0.2088 & 0.1243 & 0.5384 & 0.6850 & 0.0153 & 0.4478 & 0.2808 & 0.4799 & 0.6201 & 0.0289 & 0.2580 & 0.1658 & 0.4923 & 0.6312 & 0.0158 \\
 & \texttt{SimVP} & 0.2215 & 0.1380 & 0.5038 & 0.6514 & 0.0163 & 0.4555 & 0.3090 & 0.5442 & 0.6862 & 0.0270 & 0.2625 & 0.1802 & 0.5487 & 0.6881 & 0.0155 \\
 & \texttt{AFNO} & 0.2155 & 0.1289 & 0.5915 & 0.7322 & 0.0138 & 0.4670 & 0.2940 & 0.5536 & 0.6945 & 0.0249 & 0.2696 & 0.1621 & 0.5472 & 0.6865 & 0.0145 \\
 & \texttt{TimeXer} & 0.2143 & 0.1327 & 0.6274 & 0.7620 & 0.0128 & 0.4954 & 0.3070 & \underline{\blue{0.5838}} & \underline{\blue{0.7216}} & 0.0240 & 0.2473 & 0.1533 & 0.5934 & 0.7289 & 0.0133 \\
 & \texttt{DiffCast} & 0.2678 & 0.1449 & 0.6146 & 0.7523 & \underline{\blue{0.0107}} & 0.5641 & 0.3084 & 0.5597 & 0.7014 & 0.0216 & 0.2885 & 0.1629 & 0.5790 & 0.7183 & 0.0116 \\
 & \texttt{FlowCast} & 0.2241 & 0.1108 & 0.5994 & 0.7389 & 0.0112 & 0.5185 & 0.2636 & 0.4833 & 0.6261 & 0.0264 & \textbf{\red{0.2156}} & \textbf{\red{0.1016}} & 0.5986 & 0.7301 & \textbf{\red{0.0111}} \\
\cline{2-17}
 & \method-o & \textbf{\red{0.1777}} & \textbf{\red{0.0823}} & \textbf{\red{0.6417}} & \textbf{\red{0.7728}} & \underline{\blue{0.0107}} & \textbf{\red{0.3933}} & \underline{\blue{0.1876}} & \textbf{\red{0.6071}} & \textbf{\red{0.7408}} & \underline{\blue{0.0198}} & \underline{\blue{0.2175}} & 0.1061 & \textbf{\red{0.6066}} & \textbf{\red{0.7385}} & 0.0118 \\
 & \method-s & \underline{\blue{0.1794}} & \underline{\blue{0.0842}} & \underline{\blue{0.6404}} & \underline{\blue{0.7718}} & \textbf{\red{0.0100}} & \underline{\blue{0.3955}} & \textbf{\red{0.1851}} & \textbf{\red{0.6071}} & \textbf{\red{0.7408}} & \textbf{\red{0.0190}} & 0.2196 & \underline{\blue{0.1035}} & \underline{\blue{0.6050}} & \underline{\blue{0.7371}} & \underline{\blue{0.0112}} \\
\bottomrule[1pt]
\end{tabular}
}
\end{table*}

\subsubsection{Mean Wave Period and Direction (\texttt{GFS})}\label{app.results.mwpd_gfs}
Table~\ref{tab.exp.gfs.mwpd} reports CRPS for the auxiliary wave
variables (mean wave period, mean wave direction) on \texttt{GFS}.
\method{} substantially improves over \texttt{GFS-WW3} for both
channels --- e.g., overall CRPS-MWP drops from $1.23$\,s to $0.25$\,s
on the Gulf of Mexico. The California region is more competitive: \texttt{FlowCast} edges
\method{} on long-term MWP, consistent with the long-fetch swell
regime where \texttt{GFS-WW3} is already comparatively strong.
\begin{table*}[!t]
\centering
\caption{Mean-wave-period (MWP) and mean-wave-direction (MWD) CRPS on \texttt{GFS}'s dataset. \textbf{\red{Red}} (\underline{\blue{Blue}}) values indicate the best (second-best) per column. 
Gaussian CRPS for ensembles. CRPS reduces to MAE for non-probabilistic methods.
}\label{tab.exp.gfs.mwpd}
\vspace{0.2cm}
\tiny
\setlength{\tabcolsep}{2.5pt}{
\begin{tabular}{l|l|cc|cc|cc}
\toprule[1pt]
  &   & \multicolumn{2}{c}{\textbf{Gulf of Mexico}} & \multicolumn{2}{c}{\textbf{North Sea}} & \multicolumn{2}{c}{{\textbf{California Coast}}} \\
\cmidrule(lr){3-4} \cmidrule(lr){5-6} \cmidrule(lr){7-8}
  & \textbf{Model} & CRPS-MWP (s) & CRPS-MWD ($^\circ$) & CRPS-MWP (s) & CRPS-MWD ($^\circ$) & CRPS-MWP (s) & CRPS-MWD ($^\circ$) \\
\midrule
\multirow{11}{*}{\rotatebox{90}{Short-Term}}
 & \texttt{GFS-WW3} & 1.1789 & 18.1443 & 2.1466 & 23.6186 & 3.1095 & 24.5094 \\
 & \texttt{U-Net} & 0.3038 & 9.6514 & 0.5430 & 15.6090 & 0.7313 & 14.2569 \\
 & \texttt{EarthFormer} & 0.3223 & 10.1632 & 0.4402 & 11.7093 & \underline{\blue{0.5345}} & \underline{\blue{9.9825}} \\
 & \texttt{EarthFarseer} & 0.3196 & 10.7905 & 0.4456 & 11.5879 & 0.5360 & 10.3003 \\
 & \texttt{SimVP} & 0.3373 & 12.6680 & 0.4635 & 12.7184 & 0.5787 & 11.2427 \\
 & \texttt{AFNO} & 0.3150 & 10.2438 & 0.4403 & 11.6142 & 0.5559 & \textbf{\red{9.6755}} \\
 & \texttt{TimeXer} & 0.4500 & 16.1298 & 0.7732 & 20.5555 & 0.8732 & 14.0816 \\
 & \texttt{DiffCast} & 0.4403 & 11.5434 & 0.8880 & 13.8405 & 2.0067 & 17.2921 \\
 & \texttt{FlowCast} & 0.3789 & 15.2576 & 0.3947 & 16.7287 & \textbf{\red{0.4388}} & 12.3990 \\
\cline{2-8}
 & \method-o & \underline{\blue{0.2000}} & \textbf{\red{8.7965}} & \underline{\blue{0.3033}} & \textbf{\red{9.9402}} & 0.7044 & 10.7848 \\
 & \method-s & \textbf{\red{0.1981}} & \underline{\blue{8.8533}} & \textbf{\red{0.3018}} & \underline{\blue{9.9519}} & 0.6726 & 10.6976 \\
\midrule
\multirow{11}{*}{\rotatebox{90}{Mid-Term}}
 & \texttt{GFS-WW3} & 1.2186 & 19.5821 & 2.0705 & 25.4573 & 3.0214 & 26.0714 \\
 & \texttt{U-Net} & 0.3154 & 11.2001 & 0.5487 & 17.4141 & 0.7224 & 14.9428 \\
 & \texttt{EarthFormer} & 0.3237 & 11.2796 & 0.4722 & 13.4342 & 0.5392 & 10.5984 \\
 & \texttt{EarthFarseer} & 0.3254 & 12.2747 & 0.4737 & 13.7686 & \underline{\blue{0.5240}} & \textbf{\red{10.1840}} \\
 & \texttt{SimVP} & 0.3475 & 13.9634 & 0.4986 & 14.1441 & 0.5520 & 11.3051 \\
 & \texttt{AFNO} & 0.3303 & 11.7190 & 0.4681 & 13.9231 & 0.5441 & \underline{\blue{10.5546}} \\
 & \texttt{TimeXer} & 0.4559 & 17.3956 & 0.7616 & 22.1130 & 0.8191 & 14.7116 \\
 & \texttt{DiffCast} & 0.4584 & 13.1540 & 0.8543 & 16.2941 & 1.9572 & 19.0256 \\
 & \texttt{FlowCast} & 0.3706 & 16.4341 & 0.4014 & 17.1562 & \textbf{\red{0.4162}} & 12.5427 \\
\cline{2-8}
 & \method-o & \underline{\blue{0.2269}} & \textbf{\red{10.4695}} & \underline{\blue{0.3167}} & \underline{\blue{12.5428}} & 0.7172 & 12.5543 \\
 & \method-s & \textbf{\red{0.2205}} & \underline{\blue{10.5735}} & \textbf{\red{0.3164}} & \textbf{\red{12.5278}} & 0.6610 & 12.5271 \\
\midrule
\multirow{11}{*}{\rotatebox{90}{Long-Term}}
 & \texttt{GFS-WW3} & 1.2491 & 24.9397 & 2.0496 & 33.7038 & 3.1956 & 30.2076 \\
 & \texttt{U-Net} & 0.3946 & 17.5566 & 0.7026 & 27.8330 & 0.8941 & 17.6861 \\
 & \texttt{EarthFormer} & 0.3983 & 17.6914 & 0.6252 & \textbf{\red{24.2660}} & \underline{\blue{0.6941}} & \underline{\blue{13.9438}} \\
 & \texttt{EarthFarseer} & 0.3941 & 18.4182 & 0.6323 & 24.8971 & 0.7004 & \textbf{\red{13.9104}} \\
 & \texttt{SimVP} & 0.4305 & 19.9169 & 0.6431 & 25.4706 & 0.7044 & 14.3971 \\
 & \texttt{AFNO} & 0.4035 & 18.0726 & 0.6339 & 25.0677 & 0.7070 & 14.2969 \\
 & \texttt{TimeXer} & 0.5172 & 22.8283 & 0.8762 & 31.0028 & 0.9850 & 18.1178 \\
 & \texttt{DiffCast} & 0.5265 & 19.2589 & 0.8998 & 26.7177 & 2.2066 & 23.0010 \\
 & \texttt{FlowCast} & 0.4074 & 21.5450 & 0.5289 & 27.6177 & \textbf{\red{0.5695}} & 15.9678 \\
\cline{2-8}
 & \method-o & \underline{\blue{0.3106}} & \textbf{\red{17.3444}} & \underline{\blue{0.4756}} & 24.3245 & 0.8758 & 15.8839 \\
 & \method-s & \textbf{\red{0.2936}} & \underline{\blue{17.4844}} & \textbf{\red{0.4560}} & \underline{\blue{24.2744}} & 0.8050 & 15.9087 \\
\midrule
\multirow{11}{*}{\rotatebox{90}{Overall}}
 & \texttt{GFS-WW3} & 1.2260 & 21.5010 & 2.0677 & 29.2188 & 3.1021 & 27.4883 \\
 & \texttt{U-Net} & 0.3445 & 13.4756 & 0.6217 & 22.2269 & 0.7895 & 15.9149 \\
 & \texttt{EarthFormer} & 0.3522 & 13.6378 & 0.5423 & 18.4381 & 0.5978 & \underline{\blue{11.8002}} \\
 & \texttt{EarthFarseer} & 0.3510 & 14.4893 & 0.5474 & 18.8612 & \underline{\blue{0.5940}} & \textbf{\red{11.6156}} \\
 & \texttt{SimVP} & 0.3782 & 16.1239 & 0.5645 & 19.4154 & 0.6140 & 12.4824 \\
 & \texttt{AFNO} & 0.3567 & 14.0138 & 0.5447 & 19.0343 & 0.6087 & 11.8850 \\
 & \texttt{TimeXer} & 0.4790 & 19.3684 & 0.8173 & 26.2254 & 0.8899 & 15.9263 \\
 & \texttt{DiffCast} & 0.4825 & 15.3532 & 0.8787 & 21.0368 & 2.0611 & 20.3509 \\
 & \texttt{FlowCast} & 0.3859 & 18.2891 & 0.4616 & 22.0955 & \textbf{\red{0.4779}} & 13.8369 \\
\cline{2-8}
 & \method-o & \underline{\blue{0.2560}} & \textbf{\red{12.9524}} & \underline{\blue{0.3913}} & \textbf{\red{17.9221}} & 0.7777 & 13.6248 \\
 & \method-s & \textbf{\red{0.2474}} & \underline{\blue{13.1022}} & \textbf{\red{0.3829}} & \underline{\blue{17.9533}} & 0.7227 & 13.5947 \\
\bottomrule[1pt]
\end{tabular}
}
\end{table*}

\subsubsection{Mean Wave Period and Direction (\texttt{IFS})}\label{app.results.mwpd_ifs}

We observed that \texttt{IFS-WAM} is systematically better than
\texttt{GFS-WW3} on MWP and MWD, narrowing the gap to the
post-processed output. \method{} still wins on most cells; the
exception is California long-term MWP, where \texttt{FlowCast} edges
ahead. We suspect this is due to the Pacific long-fetch swell regime,
in which much of the wave signal arrives from outside the cropped
domain, limiting the contribution of \method's in-domain wind--wave
coupling.
\begin{table*}[!t]
\centering
\caption{Mean-wave-period (MWP) and mean-wave-direction (MWD) CRPS on \texttt{IFS}'s dataset. \textbf{\red{Red}} (\underline{\blue{Blue}}) values indicate the best (second-best) per column. 
Gaussian CRPS for ensembles. CRPS reduces to MAE for non-probabilistic methods.
}\label{tab.exp.ifs.mwpd}
\vspace{0.2cm}
\tiny
\setlength{\tabcolsep}{2.5pt}{
\begin{tabular}{l|l|cc|cc|cc}
\toprule[1pt]
  &   & \multicolumn{2}{c}{\textbf{Gulf of Mexico}} & \multicolumn{2}{c}{\textbf{North Sea}} & \multicolumn{2}{c}{{\textbf{California Coast}}} \\
\cmidrule(lr){3-4} \cmidrule(lr){5-6} \cmidrule(lr){7-8}
  & \textbf{Model} & CRPS-MWP (s) & CRPS-MWD ($^\circ$) & CRPS-MWP (s) & CRPS-MWD ($^\circ$) & CRPS-MWP (s) & CRPS-MWD ($^\circ$) \\
\midrule
\multirow{11}{*}{\rotatebox{90}{Short-Term}}
 & \texttt{IFS-WAM} & 0.4825 & 10.3872 & 0.5160 & 10.7447 & 0.8127 & 6.9564 \\
 & \texttt{U-Net} & 0.2638 & 8.9300 & 0.3395 & 8.5967 & 0.3518 & 7.4276 \\
 & \texttt{EarthFormer} & 0.2891 & 9.7657 & 0.3694 & 9.7133 & 0.3569 & 7.8750 \\
 & \texttt{EarthFarseer} & 0.2586 & 8.8383 & 0.3528 & 8.6533 & 0.3606 & 7.4498 \\
 & \texttt{SimVP} & 0.3209 & 10.5803 & 0.4212 & 12.4824 & 0.3845 & 9.2524 \\
 & \texttt{AFNO} & 0.2754 & 9.5769 & 0.3469 & 8.9027 & 0.3434 & 8.0315 \\
 & \texttt{TimeXer} & 0.3187 & 10.5029 & 0.4130 & 11.2852 & 0.4038 & 8.6414 \\
 & \texttt{DiffCast} & 0.3269 & 11.2017 & 0.4425 & 11.7581 & 0.3239 & 7.9811 \\
 & \texttt{FlowCast} & 0.2456 & 11.9952 & 0.3593 & 13.5704 & \textbf{\red{0.2214}} & 6.2442 \\
\cline{2-8}
 & \method-o & \textbf{\red{0.1643}} & \textbf{\red{7.7254}} & \underline{\blue{0.1973}} & \textbf{\red{7.1847}} & 0.2457 & \textbf{\red{6.0555}} \\
 & \method-s & \underline{\blue{0.1720}} & \underline{\blue{7.9537}} & \textbf{\red{0.1959}} & \underline{\blue{7.1848}} & \underline{\blue{0.2407}} & \underline{\blue{6.0717}} \\
\midrule
\multirow{11}{*}{\rotatebox{90}{Mid-Term}}
 & \texttt{IFS-WAM} & 0.4997 & 12.1454 & 0.5381 & 12.6323 & 0.8108 & 7.8890 \\
 & \texttt{U-Net} & 0.2804 & 10.3052 & 0.3699 & 10.1466 & 0.3709 & 7.7827 \\
 & \texttt{EarthFormer} & 0.3019 & 10.7854 & 0.3984 & 11.0200 & 0.3873 & 7.9654 \\
 & \texttt{EarthFarseer} & 0.2877 & 10.9556 & 0.3907 & 10.5443 & 0.4077 & 8.3241 \\
 & \texttt{SimVP} & 0.3388 & 11.8804 & 0.4472 & 13.6954 & 0.3966 & 9.4051 \\
 & \texttt{AFNO} & 0.2913 & 10.8475 & 0.3804 & 10.5866 & 0.3717 & 8.5253 \\
 & \texttt{TimeXer} & 0.3438 & 11.9428 & 0.4574 & 12.8496 & 0.4320 & 9.2887 \\
 & \texttt{DiffCast} & 0.3491 & 12.7037 & 0.4195 & 13.0103 & 0.3935 & 8.6919 \\
 & \texttt{FlowCast} & 0.2536 & 12.7814 & 0.3702 & 14.4779 & \textbf{\red{0.2417}} & \textbf{\red{6.6746}} \\
\cline{2-8}
 & \method-o & \underline{\blue{0.1880}} & \textbf{\red{9.3481}} & \underline{\blue{0.2355}} & \textbf{\red{8.9935}} & 0.2769 & \underline{\blue{6.8038}} \\
 & \method-s & \textbf{\red{0.1879}} & \underline{\blue{9.5560}} & \textbf{\red{0.2288}} & \underline{\blue{9.0132}} & \underline{\blue{0.2640}} & 6.8207 \\
\midrule
\multirow{11}{*}{\rotatebox{90}{Long-Term}}
 & \texttt{IFS-WAM} & 0.6153 & 18.7502 & 0.7651 & 23.6216 & 0.8552 & 12.1642 \\
 & \texttt{U-Net} & 0.3785 & 16.8379 & 0.5937 & 21.3694 & 0.5165 & 11.8668 \\
 & \texttt{EarthFormer} & 0.3631 & 16.2652 & 0.5705 & 20.7036 & 0.5105 & 11.1356 \\
 & \texttt{EarthFarseer} & 0.3652 & 16.7016 & 0.5728 & 21.3109 & 0.5271 & 11.6492 \\
 & \texttt{SimVP} & 0.4010 & 17.2658 & 0.6121 & 21.7407 & 0.5029 & 12.0682 \\
 & \texttt{AFNO} & 0.3994 & 17.2320 & 0.6262 & 21.8795 & 0.5197 & 12.3993 \\
 & \texttt{TimeXer} & 0.4186 & 17.2684 & 0.6342 & 21.2561 & 0.5408 & 12.1015 \\
 & \texttt{DiffCast} & 0.4023 & 18.7721 & 0.5674 & 22.8323 & 0.4572 & 12.4695 \\
 & \texttt{FlowCast} & 0.3037 & 18.2279 & 0.4972 & 23.3455 & \textbf{\red{0.3472}} & \textbf{\red{10.0059}} \\
\cline{2-8}
 & \method-o & \underline{\blue{0.2634}} & \textbf{\red{15.0289}} & \underline{\blue{0.4262}} & \textbf{\red{18.8640}} & 0.4099 & \underline{\blue{10.2840}} \\
 & \method-s & \textbf{\red{0.2485}} & \underline{\blue{15.1735}} & \textbf{\red{0.3967}} & \underline{\blue{19.0302}} & \underline{\blue{0.3650}} & 10.3958 \\
\midrule
\multirow{11}{*}{\rotatebox{90}{Overall}}
 & \texttt{IFS-WAM} & 0.5398 & 14.3401 & 0.6175 & 16.3982 & 0.8268 & 9.3148 \\
 & \texttt{U-Net} & 0.3141 & 12.5204 & 0.4475 & 14.0368 & 0.4210 & 9.2094 \\
 & \texttt{EarthFormer} & 0.3227 & 12.6672 & 0.4579 & 14.3893 & 0.4284 & 9.1025 \\
 & \texttt{EarthFarseer} & 0.3129 & 12.8115 & 0.4525 & 14.2397 & 0.4454 & 9.4235 \\
 & \texttt{SimVP} & 0.3597 & 13.7016 & 0.5046 & 16.4859 & 0.4339 & 10.3425 \\
 & \texttt{AFNO} & 0.3287 & 13.0210 & 0.4656 & 14.4858 & 0.4215 & 9.8577 \\
 & \texttt{TimeXer} & 0.3683 & 13.7203 & 0.5166 & 15.7329 & 0.4682 & 10.2228 \\
 & \texttt{DiffCast} & 0.3654 & 14.7463 & 0.4762 & 16.4192 & 0.4069 & 9.9698 \\
 & \texttt{FlowCast} & 0.2711 & 14.6825 & 0.4152 & 17.5874 & \textbf{\red{0.2777}} & \textbf{\red{7.8263}} \\
\cline{2-8}
 & \method-o & \underline{\blue{0.2128}} & \textbf{\red{11.2305}} & \underline{\blue{0.3006}} & \textbf{\red{12.3769}} & 0.3212 & \underline{\blue{7.9698}} \\
 & \method-s & \textbf{\red{0.2082}} & \underline{\blue{11.4180}} & \textbf{\red{0.2894}} & \underline{\blue{12.4223}} & \underline{\blue{0.2969}} & 8.0056 \\
\bottomrule[1pt]
\end{tabular}
}
\end{table*}
\subsubsection{SDE Noise Sweep}\label{app.results.sweep}
\begin{figure}[!t]
\centering
\includegraphics[width=\textwidth]{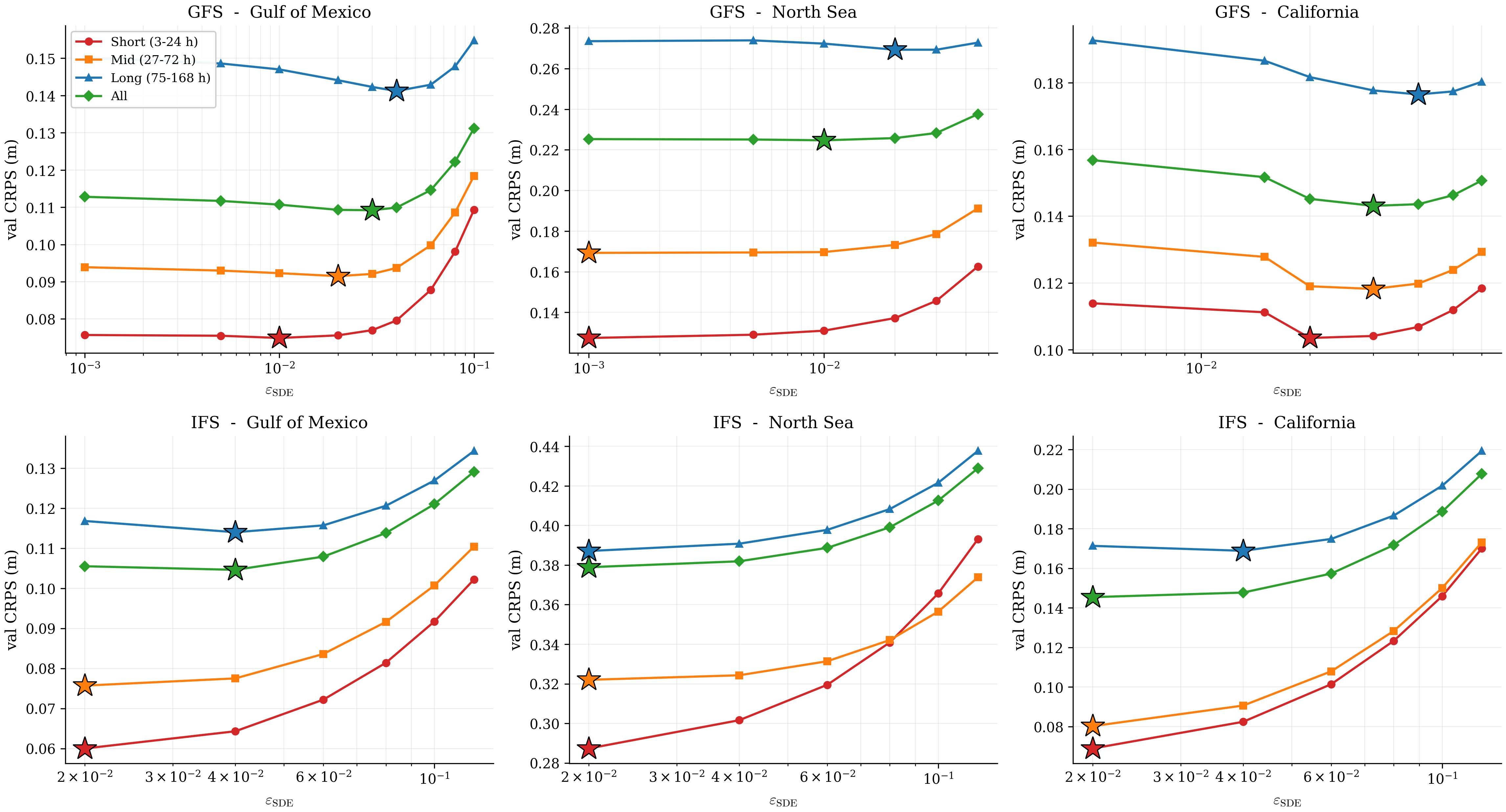}
\caption{SDE-noise sweep on the validation set. Each panel shows
validation CRPS as a function of $\varepsilon_{\text{SDE}}$,
with separate curves for Short-Term, Mid-Term, Long-Term,
and Overall. Rows correspond to forecast systems
(\texttt{GFS}, \texttt{IFS}), and columns correspond to regions
(Gulf of Mexico, North Sea, California).
Stars mark the minimum of each curve.}
\label{fig:sde_sweep}
\end{figure}

We observed three patterns. On GFS, the curves are flat over a wide
$\varepsilon_{\text{SDE}}$ range --- Gulf-of-Mexico All-band CRPS
varies by less than $4\%$ across $[10^{-3}, 3{\times}10^{-2}]$, and
North Sea is nearly invariant across the entire range. The per-band
optima migrate rightward as lead time grows (Short $\to$ Mid $\to$
Long), consistent with longer leads benefitting from larger path
noise. IFS curves are monotone-increasing, with the smallest swept
value winning in nearly every band, suggesting the IFS-trained model
is already well-calibrated at the prior-draw level. 
The per-band validation optima $\varepsilon^\star$ are shown in
Fig.~\ref{fig:sde_sweep}, though
the flat GFS curves indicate $\varepsilon_{\text{SDE}}\approx 2{\times}
10^{-2}$ would yield comparable performance.

\subsection{Implementation Details}\label{app.impl_details}

\paragraph{Architecture.}
\method{} uses a two-head \texttt{DiT}~\citep{peebles2023dit}: hidden
$D{=}256$, depth $6$, $8$ attention heads, head dim $32$, MLP ratio
$4$, no dropout. Wave grid is tokenized per-pixel; 
atmospheric tokens are
2$\times$2 down-projected before cross-attention. Inputs are
$C_{\text{in}}{=}5$ (4 wave channels + ocean mask), outputs
$C_{\text{out}}{=}4$ (SWH, MWP, MWD-sin, MWD-cos), 
atmospheric channels
$C_{\text{atm}}{=}11$, trajectory length $T{=}24$ (3-h cadence,
$\ell_{\max}{=}168$\,h). $T_l$ is injected via sinusoidal embedding
concatenated with the $\tau$ embedding through a 2-layer \texttt{SiLU MLP} and
\texttt{AdaLN-Zero} modulation. The two heads ($\hat b, \hat\eta$) share the
trunk and split via separate \texttt{AdaLN-Zero} final projections (zero-init
modulators, Xavier-init output projections); a 1$\times$1 residual
$\mathcal{X}\!\to\!\hat b$ skip is added on the velocity head only.
The ocean mask is applied multiplicatively to both heads' outputs.
Total parameters: $\approx$13.69\,M.

\begin{figure}[!t]
\centering
\includegraphics[width=0.5\linewidth]{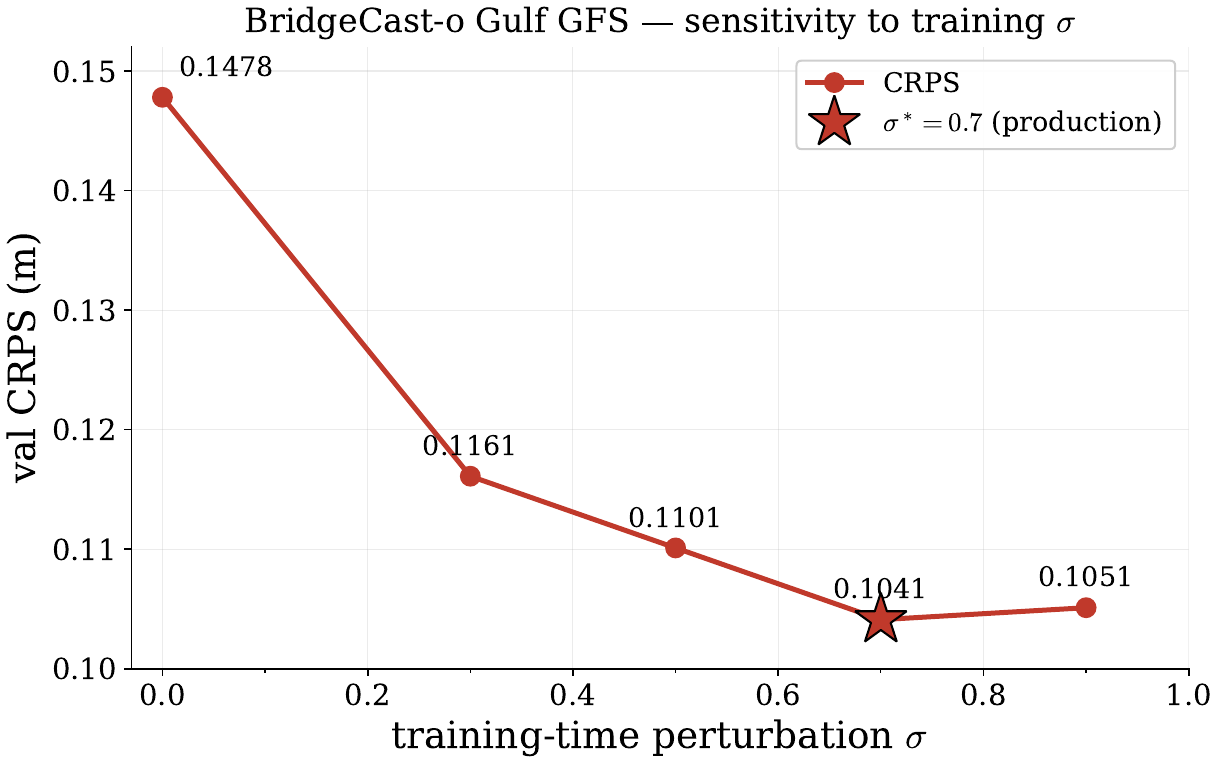}
\caption{Sensitivity to the training-time perturbation scale
$\sigma$ for \method-o on the \texttt{GFS} Gulf of Mexico validation set.}
\label{fig:sigma_sweep}
\end{figure}

\paragraph{Training.}
AdamW (PyTorch defaults for $(\beta_1, \beta_2, \epsilon)$), weight
decay $10^{-4}$, peak LR $3\times 10^{-4}$ with 5-epoch linear warm-up
followed by cosine decay to $10^{-6}$. Up to 100 epochs with patience
20 on validation SWH-RMSE. Per-GPU batch size 8, gradient accumulation
4 (effective batch 32), gradient clipping at norm 1.0. EMA on all
parameters with decay 0.999. Mixed precision (\verb|torch.cuda.amp|).

\paragraph{Loss.}
Channel weights $w = [5, 4, 1, 1]$ on (SWH, MWP, MWD-sin, MWD-cos);
ocean mask multiplied inside the squared error and used as the
denominator for normalization. Source noise scale $\sigma = 0.7$,
denoiser-loss weight $\lambda_\eta = 1$.
SI time $\tau \sim \mathcal{U}(0,1)$ is sampled per minibatch element;
forecast lead $T_l$ comes from the data sample directly
(no resampling). Antithetic noise pairing
$(\boldsymbol{\epsilon}, -\boldsymbol{\epsilon})$
is used at every training step.

\paragraph{Inference.}
Euler--Maruyama with $S=10$ steps from $\tau=0$ to $\tau=1$,
evaluating the model at $\tau_i=i/S$ for $i=0,\ldots,S-1$.
The stochastic increment is omitted in the final step.
The score denominator $(1-\tau_i)\sigma$ is floor-clamped
at $10^{-3}$.
The initial state is
$\hat{\mathcal{Y}}_0^{(k)}
= \mathcal{X} + \sigma\boldsymbol{\epsilon}^{(k)}$,
with $\boldsymbol{\epsilon}^{(k)}
\sim \mathcal{N}(\mathbf{0}, \mathbf{I})$
drawn fresh per ensemble member. Ensemble size $K=20$.
$\varepsilon_{\text{SDE}}=0$ recovers the probability-flow ODE
(\method-o); $\varepsilon_{\text{SDE}}>0$ gives the SDE sampler
(\method-s). Both variants generate ensembles through random
initial perturbations. For \method-s, per-region noise levels
are selected on the validation split
(Gulf/NS/Cali = $0.030/0.010/0.030$ for GFS,
$0.040/0.020/0.020$ for IFS).

\paragraph{Compute.}
Training and inference were performed on 
a server using eight NVIDIA RTX 6000 Ada GPUs. The inference times reported in Table~\ref{tab:inference_cost_grid} were measured using $S=10$ 
per ensemble unit per data sample.

\end{document}